\documentclass[
 reprint,
 superscriptaddress,
 nofootinbib,
 amsmath,amssymb,
 aps,
 prc,
 floatfix,longbibliography
]{revtex4-2}

\usepackage{braket}
\usepackage{amsmath}
\usepackage{graphicx}
\usepackage{dcolumn}
\usepackage{bm}
\usepackage{hyperref}
\hypersetup{
    colorlinks=true, 
    linkcolor=blue,
    citecolor=blue,
    urlcolor=blue 
}
\AtBeginDocument{%
}
\makeatletter
\AtBeginDocument{%
  \renewcommand{\equationautorefname}{Eq.\@autoref@insert@tagform}%
  \def\@autoref@insert@tagform~#1\null{~(#1)\null}%
}
\makeatother
\AtBeginDocument{%
}
\usepackage{siunitx}
\usepackage{physics}
\usepackage[table]{xcolor}
\usepackage{setspace}
\usepackage{booktabs}
\usepackage{color,soul}
\usepackage{diagbox}
\usepackage{isotope}
\usepackage{orcidlink}
\usepackage{comment}
\usepackage{pgfplots}
\pgfplotsset{compat=1.18}
\usepgfplotslibrary{colorbrewer} 
\pgfplotsset{
    colormap name=viridis,
}
\usepgfplotslibrary{groupplots}
\usepackage{tikz}
\usetikzlibrary{intersections}
\usetikzlibrary{decorations.pathmorphing}

\def\equationautorefname{Eq.}

\newcommand{\abinitio}{\textit{ab initio}~}
\newcommand{\AbInitio}{\textit{Ab Initio}~}
\newcommand{\NNb}{$N\bar{N}$~}

\newcommand{\myfigsref}[2]{%
  Figs.~\ref{#1} and~\ref{#2}%
}

\allowdisplaybreaks

\begin{document}

\title{Light antiproton-nucleus systems at low energies with the \abinitio NCSM/RGM method
}

\author{Alireza Dehghani\orcidlink{0000-0003-2350-1433}}
\email{Contact author: alireza.dehghani@ijclab.in2p3.fr}
\affiliation{Universit\'e Paris-Saclay, CNRS/IN2P3, IJCLab, 91405 Orsay, France}%
\author{Guillaume Hupin\orcidlink{0000-0002-4285-7411}}
\affiliation{Universit\'e Paris-Saclay, CNRS/IN2P3, IJCLab, 91405 Orsay, France}%
\author{Sofia Quaglioni\orcidlink{0000-0002-7512-605X}}
\affiliation{Lawrence Livermore National Laboratory, P.O. Box 808, L-414, Livermore, California 94551, USA}%
\author{Petr Navr\'atil\orcidlink{0000-0002-7493-5293}}
\affiliation{TRIUMF, 4004 Wesbrook Mall, Vancouver, British Columbia, V6T 2A3, Canada}%

\date{\today}
\begin{abstract}
The availability of low-energy antiproton beams at the CERN Antiproton Decelerator has renewed interest in using antimatter as a probe of nuclear structure and in forming exotic antiprotonic few-body systems. In this work, we extend the \abinitio no-core shell model combined with the resonating group method (NCSM/RGM), which was successfully applied to light-nucleus structure and reactions, to antiproton-nucleus dynamics at low energies.
The NCSM/RGM formalism is adapted to antiproton projectiles by removing the requirement of antisymmetrization under exchange of target and projectile constituents, while retaining a fully microscopic description of the nuclear target and the relative motion. We focus on the lightest systems, ${\bar p}+d$, ${\bar p}+\isotope[3]{H}$, and ${\bar p}+\isotope[3]{He}$, for which benchmarking against exact solutions of the Schrödinger equation enables stringent validation and helps disentangle methodological uncertainties---e.g., those associated with the choice of configurations included in the NCSM/RGM expansion---so that the dominant residual uncertainty can be attributed to the \NNb interaction. We compute phase shifts, scattering lengths, cross sections, antiprotonic-atom level shifts and widths, nuclear quasibound energies, and annihilation densities. We find that the hard short-range components of the meson-exchange-based \NNb interaction lead to slow convergence of the NCSM/RGM kernels expanded in a harmonic-oscillator basis, requiring exceptionally large model spaces and posing significant numerical challenges. We discuss practical strategies to mitigate these limitations and assess the impact of missing closed-channel configurations, which is a significant source of uncertainties in very light systems.
\end{abstract}

\maketitle

\section{Introduction} \label{sec:intro}
New experiments at CERN~\cite{caravita2025cern} have once again inspired the theoretical study of antiprotonic systems~\cite{lazauskas2021antiproton, Duerinck:2026otx,duerinck2023antiproton, Vorabbi:2019ciy}. In a challenging attempt, the antiProton Unstable Matter Annihilation (PUMA) experiment~\cite{puma} will study the surface properties of stable and rare isotopes using low-energy antiprotons. Antiprotonic probes offer a unique sensitivity to the tail of the nuclear density, making them a promising candidate for studying the surface phenomena, such as halo nuclei and neutron skins. By examining the products of annihilation, i.e., the charge of the outgoing pions, one can obtain the ratio of the annihilated protons to neutrons. From this ratio, which is the observable of PUMA, one can infer the abundance of neutrons or protons at the periphery. \par
To establish a theoretical framework for supporting the experiments with antiprotons, we attempt to study antiproton-nucleus systems from first principles. Although nuclear many-body methods have been able to attack a considerable portion of the nuclear chart~\cite{hergert2020guided}, \textit{ab initio} calculations for antiproton-nucleus systems are very rare. Antiprotonic systems are, in other words, relatively untouched. In this work, we move one small step towards filling this void by developing and implementing a formalism for the systematic study of antiproton-nucleus systems. Here, we focus on light systems in which comparison with exact few-body methods is possible, allowing us to assess the uncertainty of the many-body method. However, the main goal is to lay the foundation for extending calculations to systems with $A \approx 16$, which are beyond the reach of few-body methods and for which our many-body method works much better.\par
To study these systems, we employ the no-core shell model (NCSM)~\cite{Barrett:2013nh, Navr_til_2000} and the no-core shell model/resonating group method (NCSM/RGM)~\cite{Quaglioni:2008sm,quaglioni, unified}. The NCSM is a numerical method for studying the static properties of nuclei. This method is based on the expansion of a system's total wave function using an antisymmetrized harmonic-oscillator (HO) basis and can be implemented using either a Jacobi or a Slater-determinant (SD) basis. In the first case, the basis is made up of translationally invariant HO states built on Jacobi relative coordinates~\cite{Navr_til_2000}; in the second case, the basis is built from single-particle HO wave functions that are antisymmetrized using Slater determinants~\cite{navratil2000large}. The center-of-mass term of the Hamiltonian is absent in the Jacobi calculations. In the SD method, despite the presence of this term in the Hamiltonian, it is still possible to factor out the relative part of the total wave function~\cite{navratil2008ab}. The antisymmetrization of the basis states is straightforward in the SD method, while it is more involved in the Jacobi method. As a result, except for light systems, the SD method is usually preferred. \par
The NCSM/RGM, formulated by combining the NCSM with the RGM~\cite{TANG1978167}, is capable of describing both the dynamic and static properties of nuclear systems. The RGM assumes that the system is composed of two or more clusters or partitions in relative motion with respect to each other. In its most general form, the RGM considers all possible ways of grouping the particles into different partitions. The total wave function is expanded using the eigenfunctions of each cluster, and the expansion coefficients (i.e., the relative motion amplitudes) are unknown. In the NCSM/RGM, the cluster wave functions are obtained from NCSM calculations. There are two different, yet equivalent, formulations, depending on whether the Jacobi or SD version of the NCSM is used to obtain the cluster wave functions. The Jacobi NCSM/RGM can achieve large model spaces, but is limited to light nuclei. On the other hand, the SD NCSM/RGM is numerically more restricted in terms of the model space size, but can be easily generalized to heavier nuclei. In this version, it is possible to obtain the projectile wave function using either Jacobi~\cite{quaglioni} or SD NCSM~\cite{ATKINSON2025139189}. \par
In this work, we employ the Jacobi NCSM/RGM with two partitions (target and antiproton projectile) to investigate antiprotonic deuteron, triton, and ${}^3$He systems. Our use of the SD method is limited to checking consistency with the Jacobi calculations in small model spaces. For antiprotonic systems, NCSM/RGM formalism simplifies significantly owing to the removal of the target-projectile antisymmetrizer needed previously due to the presence of nucleon(s) in the projectile~\cite{quaglioni}. At the same time, the nucleon-antinucleon ($N\bar{N}$) interaction leads to very large two-body matrix elements when evaluated in an HO basis. As a result, the rate of convergence is significantly reduced, and exceptionally large model spaces are required for the NCSM/RGM potential to converge. We will also see that using such ``hard'' potentials introduces some numerical artifacts that need to be addressed to get sensible results. \par
This paper is organized as follows. In \autoref{sec:optical}, we briefly discuss the $N\bar{N}$ optical potentials. In \autoref{sec:rgm_formalism}, we introduce the NCSM/RGM formalism for an antiproton projectile. The results for light targets are discussed in \autoref{sec:results}. Finally, in \autoref{sec:conclusion}, we present a summary of this work.
\section{Optical \texorpdfstring{$N\bar{N}$}{NNbar} potential} \label{sec:optical}
One of the most prominent features of the $N\bar{N}$ system is annihilation. When an antinucleon (either $\bar{p}$ or $\bar{n}$) comes into contact with a nucleon (either $p$ or $n$), they annihilate, converting their mass primarily into mesons. In most cases, the final state of the annihilation is dominated by pions($\pi$): On the one hand, pions are light, so they are the preferred mechanism to release energy; on the other hand, heavier mesons rapidly decay, predominantly into pions. Because pions are much lighter than nucleons, mass-energy conservation allows the production of multiple pions even at threshold. As a result, several multipion final states exist \cite{orfanidis1973nucleon}, which makes it impractical to study antiproton annihilation by treating all individual channels separately.\par
Because of this impracticality, antiprotonic systems are often treated with phenomenological methods. A widely used framework is the optical potential method, adopted in this work. In this approach, one mimics the combined effect of all annihilation channels using a complex optical potential. 
These complex potentials can be written as
\begin{equation}
    V = U + W,
\label{eq:pot_optic}
\end{equation}
where $U$ is a real meson-exchange component while $W$ is a complex term responsible for the annihilation. The use of complex potentials of the form \autoref{eq:pot_optic} results in a non-Hermitian (complex symmetric) Hamiltonian. Furthermore, the scattering matrix, $S$, is no longer unitary, and, e.g., for an uncoupled channel, is given by
\begin{equation}
S = e^{2i (\delta_R+i\delta_I)} = e^{2i \delta_R} e^{-2 \delta_I}, 
\end{equation}
where $\delta_R$ and $\delta_I$ denote the real and imaginary parts of the phase shift, and $e^{-2 \delta_I}$ is known in the literature as the inelasticity parameter. \par
Similar to the nucleon-nucleon ($NN$) interaction, the $N\bar{N}$ interaction is formulated by the exchange of mesons. In fact, the meson-exchange contributions in the $NN$ and $N\bar{N}$ interactions are related by the so-called $G$-parity rule~\cite{PhysRev.76.1739,richard2020antiproton,richard2022nucleon,klempt2002antinucleon}, which relates the potentials of the same isospin. The $G$-parity operator is defined as a rotation in the isospin space followed by a charge conjugation~\cite{Ericson:1988gk}, i.e.,
\begin{equation}
    G = C e^{i \pi I_2},
\end{equation}
where $I_2$ is the projection of the isospin operator on $y$ axis and $C$ is the charge conjugation operator. The $G$-parity rule works as follows. Suppose we have a realistic $NN$ meson exchange interaction. To obtain the $N\bar{N}$ interaction from it, we first calculate the $G$-parity of the exchanged meson(s). If the $G$-parity is positive, then the contribution of the exchange of that meson to the potential is the same in $NN$ and $N\bar{N}$; otherwise, the contributions have opposite sign. Mathematically, this is expressed as
\begin{equation}
    U_{N\bar{N}}=\sum_i G^{(i)} U^{(i)}_{NN},
\end{equation}
where the summation runs over the different meson exchange contributions, and $G^{(i)}$ refers to the $G$-parity of the exchanged meson(s). \par
In Ref.~\cite{carbonell2023comparison}, strong \NNb phase shifts from several popular optical models, including the Paris~\cite{el2009paris}, Dover and Richard~\cite{dr2}, and Kohno-Weise~\cite{kohno1986proton} potentials, are compared with the partial wave analysis (PWA) from the Nijmegen group~\cite{zhou2012energy}. 
The phase shifts from the effective-field-theory-based potential from the Jülich group~\cite{dai2017antinucleon} were obtained by fitting to the Nijmegen PWA and are therefore not included in comparison. Despite the fact that these models yield very close predictions for integrated elastic, annihilation, and charge exchange cross sections, which are obtained from summation over different partial waves, they exhibit large discrepancies in the phase shifts in individual partial waves. 
These discrepancies are usually attributed to the scarcity of \NNb experimental data, which leads to \NNb potentials that are less constrained than their $NN$ counterpart. As a result, a certain degree of uncertainty due to \NNb interaction is inevitable in our few-body calculations. \par
The results of this work have been obtained using the Kohno-Weise (KW) $N\bar{N}$ potential. The real component, $U(r)$, of this potential is obtained from the Ueda $NN$ interaction~\cite{ueda1979antinucleon} by means of the $G$-parity transformation and includes $\pi$, $\rho$, $\omega$, and $\sigma$ meson exchanges as dominant contributions. At values $r \leq 1$ fm, the real potential is extrapolated by matching the one-boson exchange potential to a Woods-Saxon. The annihilation term of the KW potential, $W(r)$, is given by a state- and energy-independent Woods-Saxon potential
\begin{equation}
    W(r) =- \frac{W_0}{1+e^{\frac{r-R}{a}}},
\end{equation}
where the parameters are obtained by fitting to the $p\bar{p}$ total, elastic, and charge exchange cross sections. The values obtained from the fit are $W_0=1.2i$ GeV, $R=0.55$ fm, and $a=0.2$ fm. We note that the $N\bar{N}$ potential is obtained in the isospin basis. The isospin formalism for the $N\bar{N}$ system is discussed, e.g., in Ref.~\cite{carbonell2023comparison}. 
\section{\AbInitio NCSM/RGM for antiprotonic systems} \label{sec:rgm_formalism}
\subsection{Partitioning the Hamiltonian: Coupling $NN$ and \NNb sectors}
In this work, we focus on light nuclear systems interacting with an antiproton over a broad energy domain. Our goal is to determine, within a single consistent framework, both nuclear and atomic properties of antiproton-nucleus systems: The bound state of the nuclear constituent, possible strong quasibound states of the ${\bar p}$+nucleus, and atomic levels (shifts and widths) of the antiprotonic atoms. These observables span a range of scales---from hundreds of MeV for nuclear quasibound states down to the eV level for atomic energy shifts---posing a genuine multiscale problem that we address with the same wave-function ansatz throughout. We start from the nonrelativistic, time-independent Schr\"odinger equation for a system of $A$ particles
\begin{equation}
    H \left| \psi^{J^\pi T} \right\rangle = E \left| \psi^{J^\pi T}  \right\rangle,
\label{eq:TISE}
\end{equation}
where $J$, $\pi$, and $T$ denote the total angular momentum, parity, and isospin of the system, respectively. By introducing an appropriate set of relative coordinates, the Hamiltonian in the center-of-mass frame can be written as
\begin{equation}
    H = H_{\text{target}}^{(A-1)}+H_{\text{projectile}}^{(1)}+ H_{\text{relative}}^{(A-1,1)}.
\label{eq:h_factorization}
\end{equation}
Here, $H_{\text{target}}^{(A-1)}$ denotes the intrinsic Hamiltonian of the target nucleus. Its eigenvalues provide the energy spectrum of the nuclear constituents entering the antiprotonic system. Likewise, $H_{\text{projectile}}^{(1)}$ is the intrinsic Hamiltonian of the antinucleonic projectile, which in this work is restricted to a single antiproton (associated with the particle index $A$). For a single-particle projectile, this term vanishes; for a compound projectile, it would instead generate the energy spectrum of the projectile.\par
In \autoref{eq:h_factorization},  $H_{\text{relative}}^{(A-1,1)}$ collects the relative kinetic energy and the short-range interactions between the antiproton and the nuclear target constituents. The relative Hamiltonian can then be written as
\begin{align}
    H_{\text{relative}}^{(A-1,1)} =& \,T_{\text{rel}}\left(\vec{r}_{A-1,1}\right) \notag\\&+ \sum_{i=1}^{A-1} V_{N\bar{N}}\left(|\vec{r}_i-\vec{r}_A|,\vec{\sigma}_i,\vec{\sigma}_A,\vec{\tau}_i,\vec{\tau}_A\right)\notag\\&+\sum_{i=1}^{A-1} V_c \left(|\vec{r}_i-\vec{r}_A|,\tau_{zi},\tau_{zA}\right).
\label{eq:hrelative}
\end{align}
The first term on the right-hand side, $(T_{\text{rel}})$, is the relative kinetic term between the target and projectile,
\begin{equation}
    T_{\text{rel}}(\vec{r}) = -\frac{1}{2\mu} \frac{1}{r} \frac{d^2}{dr^2} r + \frac{\hat{L}^2}{2 \mu r^2}\,,
    \label{eq:treldef}
\end{equation}
which depends on the relative coordinate between the center-of-mass of the target and the projectile,
\begin{equation}
    \vec{r}_{A-1,1} = \frac{1}{A-1} \sum_{i=1}^{A-1} \vec{r}_i - \vec{r}_{A}.
    \label{r_a_a}
\end{equation}
The second term, $(V_{N\bar{N}})$, denotes the strong interaction between the antiproton and the target nucleons, with the vectors $\vec{\sigma}$ and $\vec{\tau}$ denoting, respectively, the particles' spin and isospin coordinates. The last term in \autoref{eq:hrelative} is the \emph{attractive} Coulomb interaction between the antiproton and the target protons, which can be written as 
\begin{equation}
    V_c = -\sum_{i=1}^{A-1}\frac{\hat{c}_i e^2}{|\vec{r}_i-\vec{r}_A|},
    \label{eq:coul_short}
\end{equation}
where $\hat{c}_i$ is the charge operator defined as $\hat{c}_i = \frac{1}{2}(1+ \tau_{zi})$ when acting on nucleons. We will refer to this term as the ``point-Coulomb'' interaction. \par
To \autoref{eq:h_factorization}, we add and subtract $\bar{V}_c(r_{A-1,1})$, which is the Coulomb interaction between the target treated as a pointlike charged nucleus and the antiproton. Different from $V_c$ in \autoref{eq:hrelative}, which acts on the individual nuclear constituents, this interaction depends on the relative distance between the antiproton and the center-of-mass of the target nucleus and is defined as
\begin{equation}
    \bar{V}_c = -\frac{Z e^2}{|\vec{r}_{A-1,1}|},
\label{eq:average_coulomb}
\end{equation}
where $Z$ is the number of target protons (see also \autoref{subsec:evaluation}). In the literature, this term is referred to as the ``average Coulomb'' potential.
\subsection{NCSM/RGM ansatz} \label{subsec:wave}
In this section, we briefly review the NCSM/RGM formalism originally introduced in Ref.~\cite{quaglioni} and highlight the specific features and simplifications that arise when applying it to ${\bar p}$-nucleus systems. We begin from the RGM ansatz, written as
\begin{align}
    & \psi^{J^\pi T}(\vec{r}_{A-1,1}) =  \sum_{\nu} \frac{u^{J^\pi T}_\nu(r_{A-1,1})}{r_{A-1,1}}  \notag\\ &\times  \left[ \left(\left| A-1 \,\, I_1^{\pi_1} \alpha_1 T_1 \right\rangle \left| \frac{1}{2} \, \frac{1}{2}\right\rangle \right)^{sT} Y_\ell \left(\hat{r}_{A-1,1}\right)  \right]^{J^\pi T}.
\label{eq:rgm_ansatz}
\end{align}
Here, $u_\nu(r_{A-1,1})$ and $Y_\ell(\hat{r}_{A-1,1})$ denote, respectively, the radial and angular parts of the unknown relative motion amplitude for channel $\nu$. The first ket on the right-hand side represents an eigenstate of the target with total angular momentum $I_1$, parity $\pi_1$, and isospin $T_1$. The additional label $\alpha_1$ enumerates the target's eigenstates (e.g., ground state and excited state). The second ket describes the projectile. For a single antinucleon projectile, only its spin and isospin degrees of freedom appear, both equal to $1/2$.
The channel spin $s$ is obtained by coupling the target and projectile angular momenta and $\ell$ denotes their relative orbital angular momentum. These combine to the total angular momentum $J$ through $\vec{J}=\vec{s}+\vec{\ell}$. The channel label $\nu$ therefore stands for the set $\nu \equiv \{I_1,\pi_1,\alpha_1,T_1,s,\ell\}$.\par
We obtain the target wave function within the NCSM using an antisymmetrized HO basis in Jacobi coordinates. As a result, the target state is fully antisymmetric under the exchange of any two target nucleons. In the NCSM/RGM ansatz, however, antisymmetry under exchanges between the target and the projectile constituents is not enforced \textit{a priori}. Therefore, when the projectile is itself composed of nucleons, one introduces an intercluster antisymmetrizer $(\hat{\mathcal{A}})$ to enforce the Pauli principle across the full $A$-body system. In the present work, the projectile is an antiproton, and no antisymmetrization is required between the target nucleons and the projectile. Consequently, the operator $\hat{\mathcal{A}}$ can be omitted.\par
Going back to \autoref{eq:rgm_ansatz}, it is convenient to insert a Dirac delta function and recast the ansatz in an explicit integral representation
\begin{align}
    &\psi^{J^\pi T}(\vec{r}_{A-1,1})= \sum_{\nu} \int d^3r \frac{u^{J^\pi T}_\nu(r)}{r} \delta^3(\vec{r}-\vec{r}_{A-1,1}) \notag\\ & \times \left[ \left(\left| A-1 \,\, I_1^{\pi_1} \alpha_1 T_1 \right\rangle \left|\frac{1}{2} \, \frac{1}{2} \right\rangle \right)^{sT} Y_\ell(\hat{r})  \right]^{J^\pi T},
\label{eq:rgm_ansatz_position_2}
\end{align}
or in a more compact notation
\begin{equation}
    \psi^{J^\pi T}(\vec{r}_{A-1,1})= \sum_{\nu} \int  dr \, r^2 \frac{u^{J^\pi T}_\nu(r)}{r}  \left| \phi_{\nu r}^{J^\pi T} \right\rangle,  
\label{eq:rgm_ansatz_integral_form}
\end{equation}
with
\begin{align}
    &\left |   \phi_{\nu r}^{J^\pi T} \right\rangle = \frac{\delta(r-r_{A-1,1})}{rr_{A-1,1}} \notag\\ & \times  \left[ \left(\left| A-1 \,\, I_1^{\pi_1} \alpha_1 T_1\right\rangle \left|\frac{1}{2} \, \frac{1}{2}\right\rangle \right)^{sT} Y_\ell(\hat{r}_{A-1,1})  \right]^{J^\pi T}.
\label{eq:phi_nu_r_delta}
\end{align}
Introducing the Dirac delta allows us to express the relative motion in a HO basis expansion. It can be written as
\begin{align}
   \delta^3\left(\vec{r}-\vec{r}_{A-1,1}\right) &=   \sum_{n\ell'm'} R_{n\ell',b}(r) Y_{\ell' m'}(\hat{r}) \notag\\&\quad \times R_{n\ell',b}(r_{A-1,1}) Y_{\ell'm'}(\hat{r}_{A-1,1}),
\label{eq:dirac_delta}
\end{align}
where $b=\frac{1}{\sqrt{\mu\omega}}$ is the HO length with $\mu$ denoting the projectile-target reduced mass and $\omega$ the HO frequency. Substituting \autoref{eq:dirac_delta} into \autoref{eq:rgm_ansatz_position_2} and carrying out the angular integration yields
\begin{equation}
    \psi^{J^\pi T}(\vec{r}_{A-1,1})= \sum_{\nu n} \int dr \, r^2  \frac{u^{J^\pi T}_\nu(r)}{r} R_{n \ell,b}(r) \left| \phi_{\nu n,b}^{J^\pi T}\right\rangle ,
\label{eq:rgm_ansatz_nl}
\end{equation}
with
\begin{align}
    &\left| \phi_{\nu n,b}^{J^\pi T}\right\rangle
    = R_{n\ell,b}(r_{A-1,1})  \notag\\ & \times \left[ \left(\left| A-1 \,\, I_1^{\pi_1} \alpha_1 T_1 \right\rangle \left|\frac{1}{2} \, \frac{1}{2}\right\rangle \right)^{sT} Y_\ell(\hat{r}_{A-1,1})  \right]^{J^\pi T} .
\label{eq:phi_nu_n}
\end{align}
By comparing \autoref{eq:rgm_ansatz_nl} with \autoref{eq:rgm_ansatz_integral_form}, one finds that 
\begin{equation}
 \left|\phi_{\nu r}^{J^\pi T}\right\rangle = \sum_n R_{n\ell,b}(r) \left|\phi_{\nu n,b}^{J^\pi T}\right\rangle.
\label{eq:phi_nu_r}
\end{equation}
We may now recast \autoref{eq:rgm_ansatz_nl} as
\begin{align}
       \psi^{J^\pi T}(\vec{r}_{A-1,1})&= \sum_{\nu n} \left(\int dr \, r^2 \frac{u^{J^\pi T}_\nu(r)}{r} R_{n\ell,b}(r) \right)  \left|\phi_{\nu n,b}^{J^\pi T}\right\rangle \notag\\&=
        \sum_{\nu n} c_{\nu n} \left|\phi_{\nu n,b}^{J^\pi T}\right\rangle.
\end{align}
Indeed, we identify the states $\ket{\phi_{\nu n,b}^{J^\pi T}}$ as the basis in which the total wave function is expanded. For this reason, we will refer to $\ket{\phi_{\nu n,b}^{J^\pi T}}$ as the NCSM/RGM basis. \par
To obtain the unknown channel expansion coefficients $u^{J^\pi T}_\nu(r_{A-1,1})$, we substitute the ansatz \autoref{eq:rgm_ansatz_integral_form} into the Schrödinger equation \autoref{eq:TISE}. Projecting the resulting equation onto $\bra{\phi^{J^\pi T}_{\nu'r'}}$ then yields
\begin{align}
    \sum_{\nu} \int dr \, r^2   & \bigg[\left\langle\phi_{\nu' r'}^{J^\pi T} \middle|H \middle| \phi_{\nu r}^{J^\pi T} \right\rangle \notag\\ &-E \left\langle \phi_{\nu' r'}^{J^\pi T} \middle| \phi_{\nu r}^{J^\pi T} \right\rangle \bigg]\frac{u^{J^\pi T}_\nu(r)}{r} = 0,
\label{eq:rgm_equation}
\end{align}
where the bracket notation implies integration over the relative coordinate $\vec{r}_{A-1,1}$. In what follows, we refer to the first term inside the square brackets, $\bra{\phi_{\nu' r'}^{J^\pi T}}H\ket{\phi_{\nu r}^{J^\pi T}}$, as the Hamiltonian kernel. The second term, $ \langle \phi_{\nu' r'}^{J^\pi T}| \phi_{\nu r}^{J^\pi T} \rangle$, is usually called the norm kernel, denoted $\mathcal{N}^{J^\pi T}_{\nu'\nu}(r',r)$ in the literature. Here, in contrast to the nucleonic case, it simply reduces to
\begin{equation}
    \left\langle\phi_{\nu' r'}^{J^\pi T} \middle| \phi_{\nu r}^{J^\pi T} \right\rangle = \delta_{\nu\nu'}\frac{\delta(r-r')}{rr'}.
\label{eq:norm_simple}
\end{equation}
Consequently, \autoref{eq:rgm_equation} becomes
\begin{align}
    \sum_{\nu} \int dr \,r^2    & \bigg[\left\langle\phi_{\nu' r'}^{J^\pi T} \middle| H \middle|\phi_{\nu r}^{J^\pi T}\right\rangle \notag\\ & -E \delta_{\nu\nu'}\frac{\delta\left(r-r'\right)}{rr'} \bigg]\frac{u^{J^\pi T}_\nu(r)}{r} = 0.
\label{eq:rgm_equation_nnbar}
\end{align}
Finally, note that the same kernels can be defined in the NCSM/RGM basis [\autoref{eq:phi_nu_n}]. This choice yields the corresponding spectral decomposition directly in this basis.
\subsection{Evaluation of the Hamiltonian kernels} \label{subsec:evaluation}
We now turn to how the individual contributions to the Hamiltonian kernel are evaluated in order to arrive at the NCSM/RGM integrodifferential equation [\autoref{eq:rgm_integ_diff_final}]. We start with the kinetic-energy contribution. Using \autoref{eq:phi_nu_r_delta}, the integration can be carried out straightforwardly, yielding
\begin{equation}
    \left\langle \phi_{\nu' r'}^{J^\pi T} \middle| T_{\ell}(\vec{r}_{A-1,1}) \middle|\phi_{\nu r}^{J^\pi T} \right \rangle = T_{\ell}(r)\delta_{\nu \nu'} \frac{\delta(r-r')}{rr'},
\end{equation}
where
\begin{equation}
    T_{\ell}(r) = -\frac{1}{2\mu} \frac{1}{r} \frac{d^2}{dr^2} r + \frac{\ell(\ell+1)}{2 \mu r^2}.
\end{equation}
The contribution of the target Hamiltonian is simply given by
\begin{equation}
\left\langle \phi_{\nu' r'}^{J^\pi T}\middle| H^{(A-1)}_{\text{target}}\middle| \phi_{\nu r}^{J^\pi T}\right \rangle  = E^{I_1^{\pi_1}T_1}_{\alpha_1} \,\delta_{\nu \nu'} \frac{\delta(r-r')}{rr'}.
\end{equation}
Using \autoref{eq:phi_nu_r}, the total contribution of the strong potential can be written in the form
\begin{align}
    V^s_{\nu'\nu}(r',r)&=\left\langle \phi_{\nu' r'}^{J^\pi T}\middle| V_s \middle| \phi_{\nu r}^{J^\pi T}\right \rangle \notag\\& = (A-1)\sum_{nn'}^{N_{\text{max}}} R_{n \ell,b}(r) R_{n' \ell',b}(r') \notag\\& \quad \times  \left\langle \phi_{\nu' n',b}^{J^\pi T}\middle| V_{A-1,A} \middle| \phi_{\nu n,b}^{J^\pi T}\right \rangle,
\label{eq:direct_def}
\end{align}
where $(A-1)$ is the number of \NNb pairs and $V_{A-1,A}$ denotes the interaction between the $(A-1)$th nucleon in the target and the antiproton, labeled $A$. The explicit expression for the strong-interaction potential kernel is derived in Appendix D of Ref.~\cite{dehghani} for the $A=3$ and $A\geq 4$ systems. As the antisymmetrizer operator $\hat{\mathcal{A}}$ is absent, so is the exchange potential kernel of nucleon-nucleus systems~\cite{quaglioni}. This reduces the complexity of the NCSM/RGM Hamiltonian kernel and makes the norm kernel [\autoref{eq:norm_simple}] trivial.\par
The truncation of the HO basis for the relative motion, denoted by $N_{\text{max}}$, is defined as
\begin{equation}
    N_{\text{max}} = \max{(2n+\ell)} = \max{(2n'+\ell')},
\end{equation}
where $n(n')$ and $\ell(\ell')$ are the quantum numbers of the radial HO wave functions in \autoref{eq:direct_def}.
As mentioned in the Introduction, the target wave functions are obtained using the NCSM, i.e., expansion in an antisymmetrized HO basis. This basis is truncated at $N^{\rm cluster}_{\text{max}}$, which denotes the maximum number of HO excitations of all target nucleons above the minimum
configuration. Furthermore, the $s$-shell target nuclei studied in this work have positive parity, while antiprotons have negative intrinsic parity. The NCSM truncation parameter $N^{\rm cluster}_{\text{max}}$ and $N_{\text{max}}$ defined above coincide for negative total-parity $\bar{p}$-nucleus states, which is the case for the $s$ and $d$ waves. For positive total-parity states, the maximum number of HO quanta is instead given by $N_{\text{max}}=N^{\rm cluster}_{\text{max}}+1$. \par 
The total contribution from the average Coulomb potentials in \autoref{eq:h_factorization} is given by
\begin{align}
  & \frac{1}{2}\left(\bar{V}_c(r)+\bar{V}_c(r')\right)\delta_{\nu \nu'} \notag\\ & \times \left(\frac{\delta(r-r')}{rr'} - \sum_n R_{n\ell}(r) R_{n\ell}(r') \right).
\label{eq:average_coul_contrib}
\end{align}
\indent With all the considerations above, \autoref{eq:rgm_equation_nnbar} can be written as
\begin{align}
    \big[T_{\ell'}(r') +\bar{V}_c(r')&+E^{I_1'^{\pi'_1}T'_1}_{\alpha^{\prime}_1}-E\big] \frac{u^{J^{\pi}T}_{\nu^{\prime}}(r^{\prime})}{r'} \notag\\&+ \sum_{\nu} \int dr \, r^2 \, W_{\nu^{\prime}\nu}(r^{\prime},r) \frac{u_{\nu}^{J^{\pi}T}(r)}{r}  = 0,
    \label{eq:rgm_integ_diff_final}
\end{align}
where $W_{\nu^{\prime}\nu}(r^{\prime},r)$ collects contributions from both the strong and Coulomb interactions between the target nucleus and the antiproton. These Coulomb contributions come from the part expanded in the HO basis in \autoref{eq:average_coul_contrib} together with the contribution of the point-Coulomb interaction [\autoref{eq:coul_short}], which is calculated using \autoref{eq:direct_def} with $V_s$ replaced by the point-Coulomb interaction in \autoref{eq:hrelative}~\cite{dehghani}.
In the asymptotic limit $r' \rightarrow \infty$ of \autoref{eq:rgm_integ_diff_final}, all Coulomb contributions embedded in the kernel vanish because they are expanded within a finite HO model space and therefore have compact support in practice. The only Coulomb interaction that survives is the average Coulomb potential that appears explicitly outside the integral, ensuring the correct Coulomb asymptotics of the relative-motion wave function.\par 
So far, we have developed the formalism in the relative coordinates. Since the target wave function is expanded using Jacobi relative coordinates, it is convenient to introduce the Jacobi-coordinate analog of \autoref{eq:phi_nu_n}. This will allow us to transform between different coordinate systems using HO brackets~\cite{kamun}. One can, e.g., define the following set of Jacobi coordinates~\cite{quaglioni}
\begin{align}
    \vec{\xi}_0 &= \frac{1}{\sqrt{A}} \sum_{i=1}^A \vec{r}_i, \label{eq:xi0} \\
    \vec{\xi}_1 &= \frac{1}{\sqrt{2}} (\vec{r}_1-\vec{r}_2),\label{eq:xi1} \\
    \vec{\xi}_k &= \sqrt{\frac{k}{k+1}} \left(\frac{1}{k}\sum_{i=1}^k \vec{r}_i - \vec{r}_{k+1}\right), \: 2 \leq k \leq A - 2,
    \label{eq:jac_typ_2} \\
    \vec{\eta}_{A-1}& = \sqrt{\frac{A-1}{A}}\left(\frac{1}{A-1} \sum_{i=1}^{A-1} \vec{r}_i - \vec{r}_{A}\right) \notag\\ &=\sqrt{\frac{A-1}{A}} \vec{r}_{A-1,1},
    \label{eq:eta}
\end{align}
with the corresponding Jacobi momenta defined analogously.
The relation between the Jacobi-coordinate and relative-coordinate NCSM/RGM bases is given by
\begin{equation}
    \left| \phi_{\nu n,b}^{J^\pi T} \right \rangle
    = \left(\sqrt{\frac{A-1}{A} }\right)^\frac{3}{2} \left|  \phi_{\nu n,b_0}^{J^\pi T}\right \rangle,
\end{equation}
with $b_0=\sqrt{\frac{1}{m\omega}}$ and
\begin{align}
    & \left|\phi_{\nu n,b_0}^{J^\pi T}\right \rangle=
    \left[ \left(\left| A-1 \,\, I_1^{\pi_1} \alpha_1 T_1\right \rangle \left|\frac{1}{2} \, \frac{1}{2}\right \rangle \right)^{sT} Y_\ell(\hat{\eta}_{A-1})  \right]^{J^\pi T} \notag\\&
    \times R_{n\ell,b_0}(\eta_{A-1})  .
\label{eq:rgm_basis_b0}
\end{align}
One can verify that
\begin{equation}
    \left\langle\phi_{\nu' n',b}^{J^\pi T} \middle| \mathcal{O}  \middle| \phi_{\nu n,b}^{J^\pi T} \right \rangle=\left\langle \phi_{\nu' n',b_0}^{J^\pi T}  \middle| \mathcal{O}  \middle| \phi_{\nu n,b_0}^{J^\pi T} \right \rangle.
\end{equation}
Therefore, we can calculate the matrix elements of observables using the Jacobi NCSM/RGM basis $\ket{\phi_{\nu n,b_0}^{J^\pi T}}$ without any change to the rest of the formalism, which is developed using the relative coordinates.
\section{Observables for Antiproton-Nucleus Systems}
\subsection{Effective range expansion and Trueman formula}\label{subsec:result_a2}
To obtain the antiproton-nucleus scattering length, we use the Coulomb-modified effective range expansion (ERE) given by~\cite{Carbonell:1992wd,carbonell2023comparison,betheelementary}
\begin{equation}
    -\frac{1}{a_{\text{sc}}}+\frac{1}{2}r_{\text{sc}}k^2 = k^{2\ell+1} G_{\ell}(\eta)\left(C_0^2(\eta) \cot{(\delta_{\text{sc}})} + 2\eta h(\eta) \right),
\label{eq:scat_coul}
\end{equation}
where $a_{\text{sc}}$, $r_{\text{sc}}$, and $\delta_{\text{sc}}$ denote the Coulomb-modified scattering length, effective range, and phase shift, respectively. Furthermore, $\eta = -1/kB$ is the Sommerfeld parameter, where $k$ is the momentum in the center-of-mass frame and $B$ the Bohr radius. The factors $h(\eta)$ and $C_0(\eta)$ are defined as
\begin{align}
    h(\eta) &= \sum_{m=1}^{\infty} \frac{\eta^2}{m(m^2 + \eta^2)} - \frac{1}{2}\ln{(\eta^2)} - C_E,  \\
    C_0(\eta) &= \left(\frac{2 \pi \eta}{\exp(2 \pi \eta) -1 }\right)^{\frac{1}{2}},
\end{align}
where $C_E = 0.577\cdots$ is the Euler constant. Furthermore, $G_{\ell}(\eta)$ is given by
\begin{align}
     G_{0}(\eta) &= 1,  \\ 
    G_{\ell}(\eta) &=\prod_{s=1}^{\ell}\left( 1+\frac{\eta^2}{s^2}\right) ,\text{ for }\ell\neq0.
\end{align}
\indent The effective range parameters can be related to the energy levels of the antiprotonic atom. These exotic atomic states are formed due to the attractive Coulomb interaction between the antiproton and the nucleus~\cite{batty}. However, due to the strong interaction between the nucleus and the antiproton, the energy of these atomic states is shifted and broadened. The relationship between the Coulomb-modified effective range parameters and the level shifts and half-widths of the antiprotonic atom is given by the Trueman formula~\cite{TRUEMAN196157}.\par
The Trueman formula is expressed as an expansion in $x_{\lambda}$ defined as
\begin{equation}
    x_{\lambda} = \frac{a^{\lambda}_{\text{sc}}}{B^{2 \ell+1}},
\end{equation}
where $\lambda=\{\ell,s,J \}$. The expression up to the second order is given by~\cite{Carbonell:1992wd}
\begin{equation}
    z_{n,\lambda}= -\frac{4}{n} x_{\lambda} \alpha_{n,\ell} \left(1 - x_{\lambda} \beta_{n,\ell} \right).
\label{eq:trueman}
\end{equation}
Here, $n$ denotes the principal quantum number of the antiprotonic atom and
\begin{equation}
    z_{n,\lambda}= \frac{E_{n,\lambda}-E_{n}^c}{E_{n}^c},
\label{eq:convention}
\end{equation}
where $E_{n}^c$ denotes the pure-Coulomb energy level, while $E_{n,\lambda}$ denotes the modified atomic energy due to the antiproton-nucleus interaction. Furthermore, we have
\begin{align}
    \alpha_{n,0}&=1,  \\
    \alpha_{n,\ell}&= \prod_{s=1}^{l} \left(\frac{1}{s^2} - \frac{1}{n^2} \right) , \text{for } \ell \neq 0, \\
    \beta_{n,0}&=2 \left[\ln{n} + \frac{1}{n} -\psi (n) \right], \\
    \beta_{n,1}&= \alpha_{n,1} \beta_{n,0} - \frac{4}{n^3},
\end{align}
where $\psi(n)$ is the digamma function defined as $\psi(n)=\Gamma^{\prime}(n)/\Gamma(n)$. In \autoref{sec:results}, we will compare the results obtained with the Trueman relation with those obtained directly from the bound state calculations.
\subsection{Annihilation densities} \label{sec:ani_density}
As mentioned in the Introduction, antiprotonic probes are sensitive to the tail of the nuclear density, meaning that they annihilate at the target surface. This statement can potentially be investigated by calculating the annihilation densities~\cite{carbonell1989protonium,lazauskas2021antiproton,duerinck2023antiproton}. The annihilation density is related to the total width as
\begin{equation}
    \Gamma^{J^{\pi}T}  = \int dr \gamma^{J^{\pi}T}_{a}(r).
\end{equation}
The annihilation density $\gamma^{J^{\pi}T}_{a}(r)$ determines how much each target-projectile relative distance $r$ contributes to the total width. Therefore, in the literature, it is customary to associate the annihilation density with the observable probability of decay at the distance $r$. Nevertheless, it should be emphasized that the annihilation density is not an observable in itself; relating it to measurable quantities is an approximation that is useful for obtaining a first, qualitative understanding.\par
To obtain $ \gamma_{a}(r)$, we first note that
\begin{equation}
    \Gamma^{J^{\pi}T} = - 2 \,  \text{Im} \left\{\left\langle \psi^{J^{\pi}T} \middle| H \middle| \psi^{J^{\pi}T} \right\rangle \right\},
\end{equation}
where $H$ is the total Hamiltonian of the system and $| \psi^{J^{\pi}T}\rangle$ denotes the atomic bound state. Using \autoref{eq:rgm_ansatz_integral_form} and \autoref{eq:phi_nu_r}, we can write it as
\begin{align}
    & \Gamma^{J^{\pi}T} = - 2 \, \text{Im} \Big\{ \sum_{\nu \nu'} \int dr \, dr' u_{\nu}(r) u^{*}_{\nu'}(r')\notag\\  &\times (A-1)\sum_{nn'}^{N_{\text{max}}} u_{n \ell}(r) u_{n'\ell'}(r')  \left\langle \phi_{\nu' n'}^{J^\pi T} \middle| V_{A,A-1} \middle | \phi_{\nu n}^{J^\pi T} \right\rangle \Big\}, 
\end{align}
where we have used the fact that the Hermitian part of the Hamiltonian does not contribute to $\Gamma$. The annihilation density is therefore given by
\begin{align}
    &\gamma_{a}^{J^{\pi}T}(r) =  - 2 (A-1)  \, \text{Im} \Big\{ \sum_{\nu \nu'} \sum_{nn'}^{N_{\text{max}}} u_{n \ell}(r) u_{\nu}(r) \notag\\ &\times  \left\langle \phi_{\nu' n'}^{J^\pi T} \middle| V_{A,A-1} \middle | \phi_{\nu n}^{J^\pi T} \right\rangle \int  dr'  u^{*}_{\nu'}(r') u_{n' \ell'}(r')   \Big\}. 
\end{align}
With the knowledge that only the imaginary part of the matrix element contributes to the width, this expression can be written as
\begin{align}
    &\gamma_{a}^{J^{\pi}T}(r) \notag\\&=  - 2 (A-1) \,  \sum_{\nu \nu'} \sum_{nn'}^{N_{\text{max}}}  \text{Im} \Big\{ \left\langle \phi_{\nu' n'}^{J^\pi T} \middle| V_{A,A-1} \middle | \phi_{\nu n}^{J^\pi T} \right\rangle \Big\} \notag\\& \times u_{n \ell}(r) \,\text{Re} \Big\{  u_{\nu}(r)  \int  dr'  u^{*}_{\nu'}(r')   u_{n' \ell'}(r')  \Big\}  . 
\end{align}
\section{NCSM/RGM Results for Light Antiprotonic Nuclei} \label{sec:results}
In this section, we present our results for the light antiprotonic systems obtained by solving \autoref{eq:rgm_integ_diff_final} using the calculable $R$-matrix method on a Lagrange mesh~\cite{descouvemont2010r,hesse1998coupled}. To assess the accuracy of the NCSM/RGM approach for antiproton-nucleus systems, we concentrate on light systems with ${}^2$H, ${}^3$H, and ${}^3$He targets.\par
We summarize the parameters and conventions used in both the calculations and the presentation of the results. Unless stated otherwise, all observables are computed with the input values listed in \autoref{tab:input_param}. These choices are made such that, within the finite $N_{\rm max}$ model space employed, the reported results are effectively insensitive to the parameters.
In \autoref{tab:input_param}, $\hbar\omega$ denotes the HO frequency used consistently in both the NCSM and NCSM/RGM calculations. For the $R$-matrix solution, we additionally specify the channel radius $a_c$ and the number of Lagrange basis functions $n_s$. In general, whenever dealing with the wave functions of the antiprotonic atoms, we need to use a larger value of $a_c$ (hence $n_s$) to account for the spatial extension of the atomic states.
%
\begin{table}
\caption{Setup parameters for the NCSM/RGM calculations of antiproton-nucleus systems presented in this work. For all observables, convergence has been systematically checked with respect to additional parameters not listed in the table; full details are given in the main text.} 
\begin{ruledtabular}
\begin{tabular}{lc}
    Parameter                           & Value  \\ \midrule
    $\hbar \omega$                      & 20 MeV \\
    Channel radius ($a_c$)              & 30 fm \\
    Number of basis functions ($n_s$)   & 100 \\
\end{tabular}
\end{ruledtabular}
\label{tab:input_param}
\end{table}
%
\par
For the target wave functions, we employ the bare two-body chiral effective field theory interaction at fourth order (N${}^3$LO) from Ref.~\cite{entem2003accurate}. The resulting ground-state energies are shown in \autoref{tab:bin_ener} and correspond to the fully converged values obtained in our Jacobi NCSM calculations. Finally, energies shown in the figures are given in the center-of-mass frame, while momenta are quoted in the laboratory frame.
%
\begin{table}
\caption{Ground-state energies (in MeV) of nuclei calculated with the NCSM starting from the bare two-body $NN$ interaction of Ref.~\cite{entem2003accurate}. The number of HO quanta is chosen to exceed the threshold required for the full convergence of the results, matching the NCSM/RGM ansatz $N_{\rm max}$ value.}
\begin{ruledtabular}
\begin{tabular}{lc}
    Target           & $E_{\text{gs}}$ (MeV)  \\ \midrule
    ${}^2\text{H}$   & $-$2.22 \\
    ${}^3\text{H}$   & $-$7.85 \\
    ${}^3\text{He}$  & $-$7.12 
\end{tabular}
\end{ruledtabular}
\label{tab:bin_ener}
\end{table}
%
\subsection{Regulating the NCSM/RGM kernels} \label{subsec:regulators}
\begin{figure}
\centering
\includegraphics[width=0.4\textwidth]{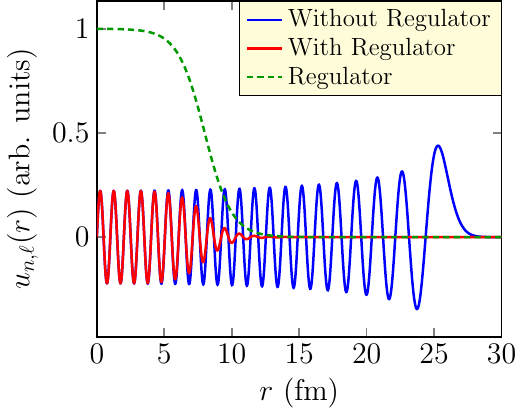}
\caption{Relative radial HO wave function ($A=2$, $n=40$, $\ell=0$, $\hbar \omega=20$ MeV) before (blue lines) and after (red) applying a Woods-Saxon regulator (green dashed line) with $r_{\text{reg}}=8$ fm.
}
\label{fig:regulator}
\end{figure}

In the limit $N_{\text{max}}\to\infty$, the NCSM/RGM potential kernel generated by the short-range $NN$ and \NNb interactions becomes confined to the interaction region: It is non-negligible only when the target and projectile overlap, and it vanishes at large intercluster separations. This behavior follows from the finite range of the underlying strong interactions. As indicated, e.g., by \autoref{eq:direct_def} for the strong-interaction part, the potential kernel $V_{\nu' \nu}(r',r)$ is built from overlaps of HO wave functions. At finite $N_{\max}$, the truncated HO basis leaves residual, almost negligible contributions stemming from the asymptotic tails of the HO functions, which can induce small spurious components in the kernel at large $r$ or $r'$. We refer to these numerical artifacts as finite-model-space effects. Their origin is illustrated in \autoref{fig:regulator}, where the blue curve shows a representative high-node HO function (often exceeding $\approx 20$ nodes in the present calculations). A regulator (green curve) may be applied so that the spatial support of the HO wave functions is confined to the interaction region. \par
%
\begin{figure}
\centering
\includegraphics[width=0.4\textwidth]{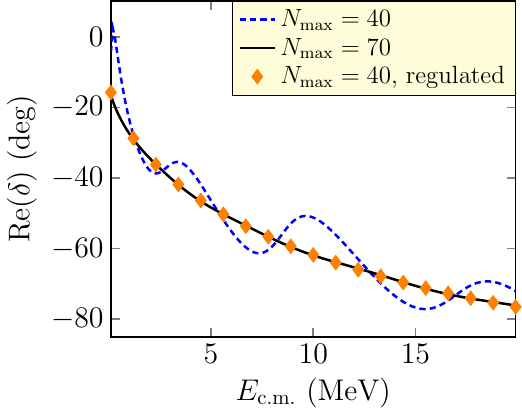}
\caption{Effect of the finite-model-space artifacts and their removal on the real part of the $s$-wave $\bar{p}$-$d$ phase shift in the ${}^2S^-_{1/2}$ channel. The calculations use the deuteron ground-state and $a_c=20$ fm. For the regulated calculation, we use $r_{\text{reg}}=10$ fm and $r_{\text{reg,\text{c}}}=5$ fm for the strong and short-range Coulomb contributions to the kernels, respectively. The blue dashed curve illustrates the oscillation of the phase shift around the physical value due to the presence of the numerical artifacts. The black curve demonstrates suppressing the artifacts by increasing the size of the model space, while the yellow diamonds demonstrate doing so using a regulator.}
\label{fig:noise}
\end{figure}

\begin{figure}
\centering
\begin{minipage}{0.15\textwidth}
\includegraphics[width=\textwidth]{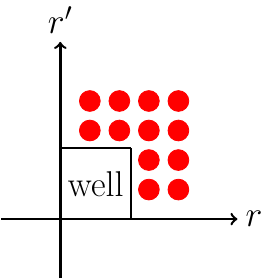}
  {\small (a) Small $N_{\text{max}}$}
\end{minipage}
\begin{minipage}{0.15\textwidth}
  \includegraphics[width=\textwidth]{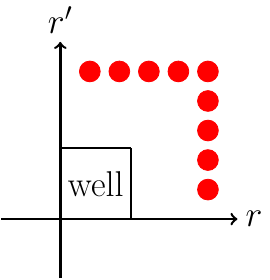}
  {\small (b) Large $N_{\text{max}}$}
\end{minipage}
\begin{minipage}{0.15\textwidth}
\includegraphics[width=\textwidth]{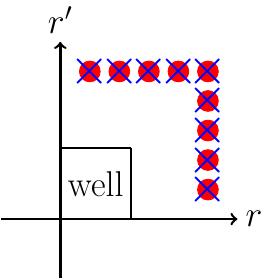}
  {\small (c) Regularization}
\end{minipage}
\caption{Schematic illustration of the naive regularization introduced to mitigate artifacts from the finite HO expansion used in the NCSM/RGM kernels. Filled circles denote the finite-model-space artifacts induced by the hard-core nature of the \NNb interaction, whereas crossed circles represent their removal to obtain the desired smooth asymptotic behavior outside the antinucleon-nucleus interaction region. This cartoon motivates the numerical results shown in \autoref{fig:noise}.
}
\label{fig:remove_noise}
\end{figure}

%
We now turn to the effect of the presence of these numerical artifacts on observables. In such a situation, the scattering solution---although numerically stable and reproducible with independent checks---can display phase shifts that oscillate around the physical value. This behavior is illustrated by the blue dashed curve in \autoref{fig:noise} for the $\bar{p}$-$d$ $s$-wave phase shifts. The numerical artifacts (hence oscillations) are progressively suppressed as the model space is enlarged—from $N_{\text{max}}=40$ (blue dashed line) to $N_{\text{max}}=70$ (black solid line). In other words, increasing $N_{\rm max}$ extends the distance over which the kernels behave as in the $N_{\text{max}}\to\infty$ limit. However, pushing $N_{\rm max}$ to very large values is not always computationally feasible, which motivates a practical solution. A practical approach is to suppress the unphysical large-distance HO tails while preserving the short-range interaction region. We propose to (i) perform calculations at sufficiently large $N_{\text{max}}$ to ensure convergence in the interior and then (ii) remove the long-range artifacts by applying a smooth regulator to the HO wave functions at large distances, leaving the potential well intact. A schematic illustration of the procedure is shown in \autoref{fig:remove_noise}. The idea can be applied to any short-range interaction represented in a truncated HO basis in order to prevent spurious long-range behavior. In this work, we adopt a Woods-Saxon-type regulator of the form
\begin{equation}
    f_s(r)= \frac{1}{1+\exp(r-r_{\text{reg}})}  
\end{equation}
for the strong-interaction part and
\begin{equation}
    f_c(r)= \frac{1}{1+\exp(r-r_{\text{reg,\text{c}}})}
\end{equation}
for the short-range Coulomb part. The effect of these regulators on the HO wave function is shown in \autoref{fig:regulator} (red line). We apply these regulators to the strong and Coulomb contributions separately and denote the regulator parameters as $r_{\text{reg}}$ and $r_{\text{reg,\text{c}}}$, respectively. For example, for the strong-interaction part, \autoref{eq:direct_def} is modified into
\begin{align}
    V^s_{\nu'\nu}(r',r) & = (A-1)\sum_{nn'}^{N_{\text{max}}} R^{\text{reg}}_{nl,b}(r) R^{\text{reg}}_{n'l',b}(r') \notag\\
    & \quad \times \left\langle \phi_{\nu' n'}^{J^\pi T} \middle| V_{A,A-1} \middle| \phi_{\nu n}^{J^\pi T} \right\rangle,
\end{align}
with $R^{\text{reg}}_{nl,b}(r) = f_s(r)R_{nl,b}(r)$. Similarly, the regulator $f_c(r)$ is applied to the Coulomb contributions depending on the HO wave functions, which include the point-Coulomb contribution together with the contribution from the second term in the right-hand side of \autoref{eq:average_coul_contrib}. \par
As an illustration, \autoref{fig:noise} shows the application of this procedure to the $\bar{p}$-$d$ phase shifts. We find that the regulated result obtained at $N_{\text{max}}=40$ closely reproduces the unregulated calculation at $N_{\text{max}}=70$, supporting the validity of the method and indicating that the regulator effectively accelerates convergence by suppressing finite-model-space artifacts. A more detailed analysis of the regulator dependence of several observables is presented in Ref.~\cite{dehghani}.
%
%
%
%
\par
We note that the finite-model-space artifacts are generally not expected for sufficiently soft interactions. In that case, the matrix elements in \autoref{eq:direct_def} decrease rapidly with increasing model-space size, so contributions involving large-$n$ and large-$n'$ HO functions are effectively suppressed because they are multiplied by near-zero amplitudes. As shown in the next sections, this is not the case for the KW \NNb potential: Its strong short-range components lead to matrix elements that remain significant over a wider range of $n$ and $n'$. Consequently, substantially large model spaces are required before the contributions in \autoref{eq:direct_def} become negligible.\par
Finally, we note that one could have possibly dealt with the finite-model-space artifacts by choosing a smaller intercluster truncation $N_{\rm max}$ than the target $N^{\rm cluster}_{\rm max}$, thereby damping the large-distance HO tails in the relative coordinate. We have not pursued this option here, because many \NNb interactions contain pronounced short-range components (partly moderated by absorption, strong too), and maintaining a consistent, sufficiently large cluster space is important to represent the interaction region accurately. Other few-body techniques designed to treat Coulomb interactions in the presence of short-range forces are likely applicable to this problem, but we have not explored them here.\par
\subsection{Antiprotonic deuteron system} \label{subsec:result_a3}
We now turn to our first set of results, which provides a benchmark against existing calculations for the $\bar{p}$-$d$ system and allows us to assess the convergence of the method. Even though corresponding results for an $\bar{n}$ projectile can be obtained straightforwardly, we do not present them here due to the limited availability of experimental data and theoretical calculations. For targets such as deuteron, the $\bar{n}$ results can be obtained from the $\bar{p}$ calculations by using the same strong-interaction kernels, and switching off all Coulomb contributions.\par
\begin{figure}
\centering
\includegraphics[width=0.4\textwidth]{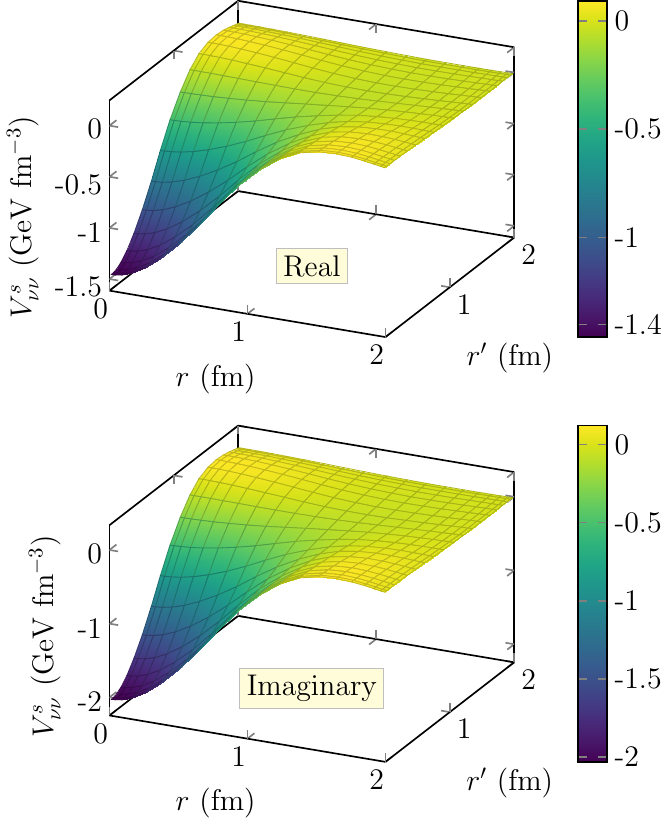}

\caption{Diagonal ($\nu=\nu'$) strong-interaction potential kernel $V^{s}_{\nu \nu}(r',r)$ for antinucleon–deuteron system in the ${}^2S_{1/2}^-$ channel, extracted from the NCSM/RGM calculation with $N_{\text{max}}=20$ and deuteron's ground state. The kernel is shown for $r,r'\le 2$ fm, beyond which it is negligible (consistent with zero on the scale of the plot).
}
\label{fig:kern_pd_-}
\end{figure}

In \autoref{fig:kern_pd_-}, we show the strong-interaction potential kernel [\autoref{eq:direct_def}] for the ${}^2S_{1/2}^-$ channel. To facilitate a direct comparison across different systems, we display the kernels at the common value of $N_{\text{max}}=20$. While this is a modest model-space size for the lightest targets, it is closer to what can be achieved for heavier systems in practical SD NCSM/RGM calculations. 
The kernels shown here include only the contribution from the target ground state. As $N_{\rm max}$ increases, the attractive pocket deepens, reflecting the large near-diagonal matrix elements of \autoref{eq:direct_def} that remain significant even for large HO radial quantum numbers $(n,n')$. The overall shape of the kernel, however, is governed by the HO basis and closely parallels what is observed in nucleon-nucleus application (see, e.g., Ref.~\cite{quaglioni}).
For deuteron---and for all systems considered here---the imaginary part is typically more attractive than the real part, consistent with strong absorption. This also helps explain why an extremely fine resolution of the intercluster interaction at the origin is not always required: The wave function is already strongly damped by the absorptive component at small distances. Although these unregulated kernels appear nearly flat at large $r$ or $r'$, finite-model-space artifacts in this region still impact scattering and atomic observables, motivating the regularization strategy discussed above.
%
%
\par
\begin{figure}
\centering
\includegraphics[width=0.4\textwidth]{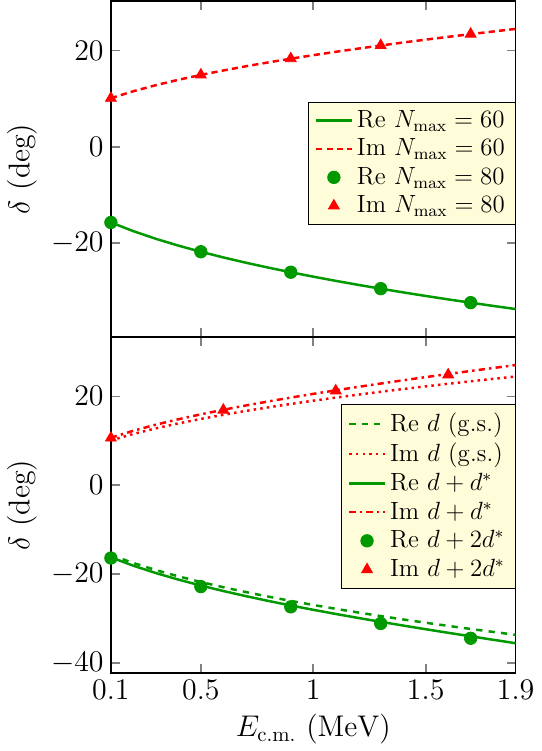}
\caption{Real (green) and imaginary (red) parts of the $s$-wave $\bar{p}$-$d$ phase shift in the ${}^2S^-_{1/2}$ channel, obtained with the regulator parameters $r_{\text{reg}}=12$ fm and $r_{\text{reg,\text{c}}}=5$ fm. Top: $N_{\text{max}}$ dependence when only the deuteron ground state is retained. Bottom: Dependence on the number of included target states in the deuteron channel at fixed $N_{\text{max}}=70$. Here $d~ {(\rm g.s.)}$ denotes the phase shift including only the deuteron ground state; $d+d^*$ includes the ground state plus the first pseudostate in the same (${}^3{S}_1$-${}^3{D}_1$) channel, and so on. At this $N_{\text{max}}$, contributions from the other two-body channels are negligible.
}
\label{fig:h2_p_phase_nmax_nstate}
\end{figure}

In \autoref{fig:h2_p_phase_nmax_nstate}, we show the $s$-wave $\bar{p}$-$d$ diagonal phase shifts in the ${}^2S^-_{1/2}$ channel, which is coupled to ${}^4D_{1/2}^-$. The top panel compares results obtained in two large model spaces, $N_{\max}=60$ (lines) and $N_{\max}=80$ (symbols). The close agreement between the two calculations demonstrates that the phase shifts can be made essentially insensitive to $N_{\max}$ at these values. \par
The light nuclei considered in this work do not have bound excited states. Nevertheless, in the NCSM, one can discretize the continuum of these nuclei and obtain solutions with eigenenergies above the ground state. We refer to these solutions as pseudostates. Incorporating these target pseudostates into the NCSM/RGM ansatz [\autoref{eq:rgm_ansatz}] leads to additional contributions for bound-state problems. On the other hand, including them in scattering calculations mimics continuum coupling without explicitly treating breakup channels. For example, for a deuteron target, one may include pseudostates in the deuteron channel, as well as $pn$ eigenstates with quantum numbers differing from those of deuteron~\cite{Hupin:2014iqa, Navratil:2011ay}. For low-energy scattering, however, energy conservation suggests that the net impact of such closed channels is typically small. This is shown in the bottom panel of \autoref{fig:h2_p_phase_nmax_nstate}, where we investigate the impact of enlarging the target basis by including up to two pseudostates in the coupled ${}^3{S}_1$-${}^3{D}_1$ deuteron channel. The comparison between different numbers of included target states indicates that the results are well converged, and that adding deuteron pseudostates produces only a minor change in the phase shifts. This is consistent with the fact that these additional channels lie above the deuteron breakup threshold and therefore remain closed over the energy range considered here. Finally, we note that due to the attractive Coulomb interaction between the target and antiproton, the $s$-wave phase shift does not vanish in the zero-energy limit~\cite{TRUEMAN196157}.\par
%
\begin{figure}
\centering
\includegraphics[width=0.39\textwidth]{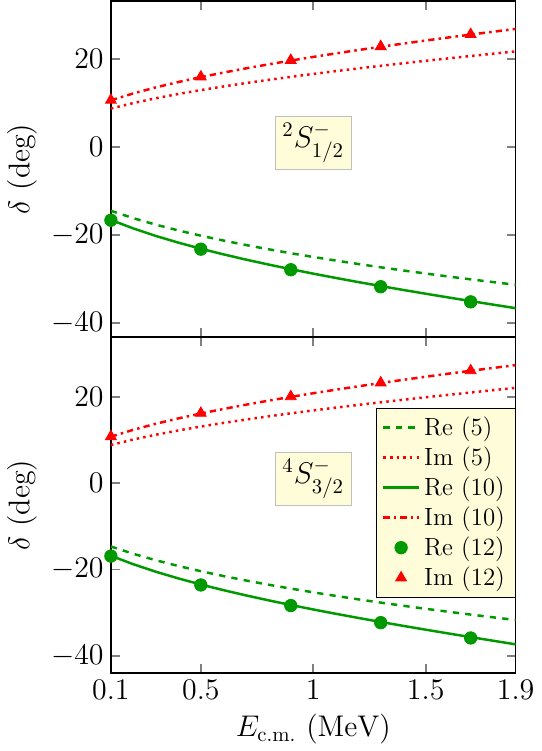}
\caption{Real (green) and imaginary (red) parts of the $s$-wave $\bar{p}$-$d$ phase shifts in the ${}^{2}S^{-}_{1/2}$ (top) and ${}^{4}S^{-}_{3/2}$ (bottom) channels. The dependence on the strong-interaction regulator radius $r_{\text{reg}}$ (in fm) is shown: $r_{\mathrm{reg}}=5$ fm (dashed), 10 fm (solid), and 12 fm (symbols), with the corresponding value indicated in parentheses in the legend. The remaining parameters are $N_{\text{max}}=60$, $r_{\text{reg,\text{c}}}=5$ fm and a deuteron model space including the ground state plus two pseudostates in the coupled ${}^3{S}_1$-${}^3{D}_1$ channel.
}
\label{fig:h2_p_phase_regdep}
\end{figure}

%
In \autoref{subsec:regulators},  we introduced two coordinate-space regulators to mitigate finite-model-space artifacts: $r_{\text{reg}}$ (strong part) and $r_{\text{reg,\text{c}}}$ (short-range Coulomb part). It is therefore essential to show the independence of observables on these regulator parameters. We have verified that, once the regulator is chosen larger than the physical interaction region (as illustrated by the kernel in \autoref{fig:kern_pd_-}), the results become effectively independent of the specific regulator values.
This behavior is exemplified in \autoref{fig:h2_p_phase_regdep}, where we compare the $s$-wave phase shifts for three choices of the strong-interaction regulator: $r_{\text{reg}}=5$ fm (dashed), 10 fm (solid), and 12 fm (symbols). A regulator radius comparable to the range of the intercluster interaction (e.g., 5 fm here) is clearly detrimental, because it suppresses a physically relevant portion of the kernel rather than only removing the finite-model-space artifact. Conversely, taking $r_{\text{reg}}$ unnecessarily large would amount to keeping the numerical artifacts in the kernel. In practice, one observes a plateau window in $r_{\text{reg}}$ where observables are stable with respect to the regulator and insensitive to other numerical parameters once convergence is reached. In \autoref{fig:h2_p_phase_regdep}, the results for $r_{\text{reg}}=10$ and 12 fm fall on top of each other, demonstrating that we are within this plateau.
As $N_{\rm max}$ increases, larger values of $r_{\text{reg}}$ can be used without compromising numerical stability---an expectation that we have checked numerically. We typically observe only a weak dependence on $r_{\text{reg,\text{c}}}$; for this reason, we do not display a separate $r_{\text{reg,\text{c}}}$ analysis in \autoref{fig:h2_p_phase_regdep}. Nevertheless, including $r_{\text{reg,\text{c}}}$ is important: In a truncated model space, the short-range Coulomb contributions can otherwise leak into---and distort---the long-range Coulomb behavior. The regulator cleanly separates these two contributions and thereby preserves the correct asymptotic Coulomb solution.\par
%
%
Similarly to the study in \autoref{fig:h2_p_phase_nmax_nstate}, \autoref{tab:scat_len_pbard} summarizes the convergence of the $s$-wave scattering length as a function of $N_{\text{max}}$ and of the number of included deuteron pseudostates. The same qualitative pattern is seen: Adding pseudostates modifies the scattering length only mildly, at the level of a few percent.
When assessing convergence with respect to  $N_{\text{max}}$ in the presence of pseudostates, one must remember that these positive-energy eigenstates depend on the NCSM truncation parameter for the target, i.e., $N^{\text{cluster}}_{\text{max}}$. Because they represent the discretized $np$ continuum, their energies shift as $N^{\text{cluster}}_{\text{max}}$ is increased; consequently, the first (or the second) pseudostates at different truncations do not have the same energies. A rigorous convergence check should therefore emphasize stability with respect to the ground-state result (which is bounded from below).
For these calculations, the regulator parameter is chosen in a plateau region. Details of this procedure are discussed in Ref.~\cite{dehghani}.\par
\begin{figure}
\centering
\includegraphics[width=\linewidth]{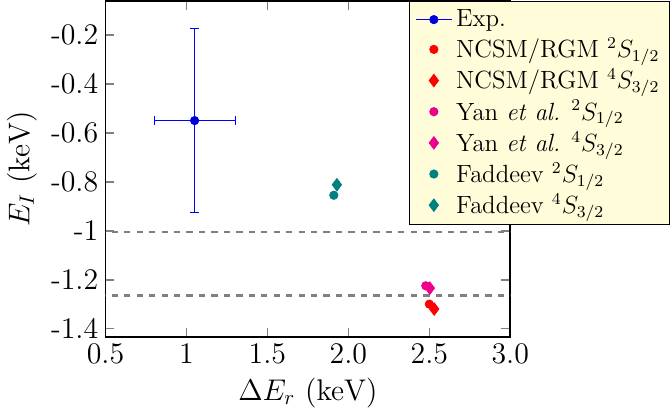}
\caption{Comparison of our $s$-wave atomic level shifts $(\Delta E_r)$ and half-widths $(E_I=-\Gamma/2)$ with other theoretical calculations and with experiment. The spin-averaged experimental value is taken from Ref.~\cite{augsburger1999measurement}. The dashed lines denote an estimation for the range of half-width by Gotta \textit{et al.}~\cite{gotta1999balmer}.
}
\label{fig:compar_shift}
\end{figure}

%
\begin{table}
\caption{$\bar{p}$-$d$ $s$-wave scattering lengths (in fm). Here $d~ {(\rm g.s.)}$ denotes a calculation including only the deuteron ground state; $d+d^*$ includes the ground state plus the first pseudostate in the same (${}^3{S}_1$-${}^3{D}_1$) channel, and so on.}
\centering
\small
\begin{ruledtabular}
\begin{tabular}{lcccc}
             & \multicolumn{2}{c}{$N_{\text{max}}=70$} & \multicolumn{2}{c}{$N_{\text{max}}=80$} \\
             & ${}^2S_{1/2}^-$  & ${}^4S_{3/2}^-$      & ${}^2S_{1/2}^-$     & ${}^4S_{3/2}^-$  \\ \midrule
    g.s.       & 1.70$-$1.09$i$    & 1.71$-$1.10$i$        & 1.70$-$1.09$i$        & 1.71$-$1.10$i$    \\
    $d+d^*$  & 1.75$-$1.15$i$     & 1.77$-$1.17$i$         & 1.74$-$1.15$i$       &   1.76$-$1.16$i$      \\
    $d+2d^*$ & 1.77$-$1.15$i$     & 1.79$-$1.17$i$         & 1.75$-$1.15$i$        &   1.77$-$1.16$i$     \\
\end{tabular} 
\end{ruledtabular}
\label{tab:scat_len_pbard}
\end{table}
%
In \autoref{tab:level_pbard_rmat}, we present the results of our bound-state $R$-matrix calculations for the level shifts and half-widths of the $s$-wave atomic states. In addition, in \autoref{tab:level_pbard}, we compute the same observables using the second-order Trueman formula [\autoref{eq:trueman}], taking as input the values of \autoref{tab:scat_len_pbard} at $N_{\text{max}}=80$. In our convention, i.e., \autoref{eq:convention}, a positive level shift indicates that the strong interaction acts repulsively. The two methods show good agreement, which is expected given the large spatial extension of the exotic atom (the ground-state radius is $\approx 43$ fm). This also shows that our scattering calculations, from which we obtain the scattering length and then the level shift, are consistent with our bound-state calculations. The comparison of our results with the spin-averaged experimental value~\cite{augsburger1999measurement}, other theoretical calculations, and an estimation by Gotta \textit{et al.}~\cite{gotta1999balmer} for the imaginary part is summarized in \autoref{fig:compar_shift}. These theoretical calculations are done by Lazauskas \textit{et al.}~\cite{lazauskas2021antiproton} using the Faddeev method and by Yan \textit{et al.}~\cite{yan2008p} using a method based on Sturmian functions. For comparison, we have selected those Faddeev results obtained using the same $NN$ interaction as ours, without including the charge exchange channel, $\bar{p}d \rightarrow \bar{n}d$. On the other hand, the results of Yan \textit{et al.} use a fully $s$-wave deuteron wave function and therefore differ from ours in that respect. \par
\begin{figure}

\centering
\begin{minipage}[t]{0.4\columnwidth}
  \centering
  \small{NCSM/RGM configuration}\par\medskip
 \includegraphics[width=0.9\textwidth]{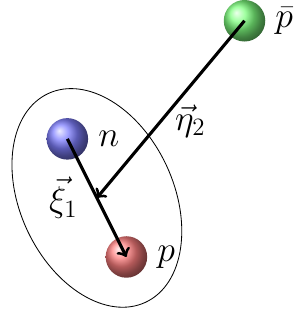}
\end{minipage}
\hfill
\begin{minipage}[t]{0.35\columnwidth}
  \centering
  \small{Other two Faddeev configurations}\par\medskip
 \includegraphics[width=0.9\textwidth]{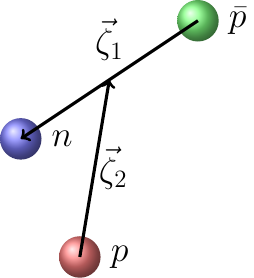}  
  \includegraphics[width=0.9\textwidth]{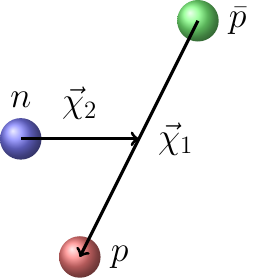}
  \vspace{1pt}
\end{minipage}

\caption{Schematic overview of the channel configurations contributing to the antiprotonic deuteron system, compared with (left) the configuration included in the present NCSM/RGM treatment. The Jacobi coordinates $\vec{\xi}_1$ and $\vec{\eta}_2$ are defined in \autoref{eq:xi1} and \autoref{eq:eta}, respectively. The remaining Jacobi coordinates are defined analogously.} 
\label{fig:rgm-configuration}
\end{figure}

Two points in this figure need comment. First, our results are in good agreement with those of Yan \textit{et al.}, whereas a discrepancy remains with the Faddeev result. This difference may stem from the fact that deuteron is only weakly bound: At short distances, the $\bar{p}pn$ dynamics may not be well represented by a $\bar{p}+d$ cluster configuration, as done here. This configuration is shown in the left panel of \autoref{fig:rgm-configuration}, together with the additional configurations present in the Faddeev method (right panel). One systematic route to improve our results is therefore to include these additional configurations, which would require specialized numerical techniques to handle eigenstates of a non-Hermitian (complex symmetric) Hamiltonian, since the \NNb interaction is absorptive. Nevertheless, we do not expect this limitation to be equally relevant for low-energy scattering from a more tightly bound nucleus, such as ${}^4$He, for which the cluster picture is more robust. \par
A second issue is the discrepancy between the Faddeev results---essentially an exact solution of the three-body Hamiltonian---and the experimental value. These Faddeev calculations are not sensitive to the $NN$ interaction used, and show only a weak sensitivity to the \NNb interactions employed. This suggests that the origin of the disagreement may lie elsewhere. A plausible explanation is the limited quality of the present two-body $N\bar{N}$ interactions, which remain insufficiently constrained by the sparse low-energy data. Otherwise, the discrepancy can point out to the presence of a $NN\bar{N}$ three-body interaction.
%
\begin{table}
\caption{$\bar{p}$-$d$ $s$-wave antiprotonic-atom level shifts and half-widths (in keV), computed with the bound-state $R$-matrix method. The quantum number $n$ denotes the principal quantum number of the antiprotonic atom. The calculations use $N_{\text{max}}=80$, $a_c=400$ fm, and $n_s=300$. The channel radius $a_c$ is chosen based on the size of the antiprotonic atom.
}
\label{tab:level_pbard_rmat}
\small
\centering
\begin{ruledtabular}
\begin{tabular}{ccccc}
        Channel   &        & $d~ {(\rm g.s.)}$  & $d+d^*$  & $d+2d^*$    \\ \midrule   
  ${}^2S_{1/2}^-$ &  $n=1$ & 2.42$-$1.25$i$   & 2.48$-$1.31$i$ & 2.50$-$1.30$i$ \\   
                  &  $n=2$ & 0.32$-$0.18$i$   & 0.32$-$0.18$i$ & 0.33$-$0.18$i$ \\
  ${}^4S_{3/2}^-$ &  $n=1$ & 2.44$-$1.26$i$   & 2.51$-$1.32$i$ & 2.53$-$1.32$i$ \\  
                  &  $n=2$ & 0.32$-$0.18$i$   & 0.33$-$0.19$i$ & 0.33$-$0.19$i$ \\
\end{tabular}
\end{ruledtabular}
\end{table}
%
%
\begin{table}
\caption{$\bar{p}$-$d$ $s$-wave antiprotonic-atom level shifts and half-widths (in keV), computed using the Trueman formula up to the second order. The calculations use the scattering lengths of \autoref{tab:scat_len_pbard} at $N_{\text{max}}=80$ as input.}
\small
\begin{ruledtabular}
\begin{tabular}{ccccc}   
       Channel    &       & $d~ {(\rm g.s.)}$  & $d+d^*$  & $d+2d^*$     \\ \midrule   
  ${}^2S_{1/2}^-$ & $n=1$ & 2.43$-$1.26$i$ & 2.49$-$1.32$i$   & 2.50$-$1.32$i$ \\ 
                  & $n=2$ & 0.32$-$0.18$i$ & 0.32$-$0.19$i$   & 0.33$-$0.19$i$ \\ 
  ${}^4S_{3/2}^-$ & $n=1$ & 2.45$-$1.27$i$ & 2.52$-$1.34$i$   & 2.53$-$1.33$i$ \\ 
                  & $n=2$ & 0.32$-$0.19$i$ & 0.33$-$0.20$i$   & 0.33$-$0.20$i$ \\  
\end{tabular}
\end{ruledtabular}
\label{tab:level_pbard}
\end{table}
%
%
%
\par
From the attractive pocket visible in \autoref{fig:kern_pd_-}, we infer that this system can also support ``nuclear quasibound states'': In contrast to the antiprotonic-atom quasibound levels, these states are generated by the strong interaction but have a large imaginary component owing to the annihilation. The corresponding energies are reported in \autoref{tab:quasi_pbard_nmax_ntar}, where we provide a convergence study analogous to that of \autoref{tab:level_pbard_rmat}. In this case, including the first deuteron pseudostate has a noticeable impact on the quasibound spectrum, whereas adding additional pseudostates produces only minor changes. Regarding the convergence with respect to $N_{\rm max}$, the results obtained with the deuteron ground state alone are already well stabilized.
These energies are obtained by diagonalizing the Hamiltonian in the NCSM/RGM basis [\autoref{eq:phi_nu_n}]. Equivalently, one may solve the NCSM/RGM coupled integrodifferential equations with the $R$-matrix method and impose bound-state boundary conditions, providing a more accurate representation of the asymptotics compared to the HO basis. We have verified that both procedures yield identical eigenvalues~\cite{dehghani}. This agreement is expected given the large model space used here (in contrast to nucleon-nucleus systems, where antisymmetrization constraints can limit the size of the basis) and provides an additional internal consistency check of our implementation and numerical workflow. \par
\begin{figure}
\centering

\includegraphics[width=0.4\textwidth]{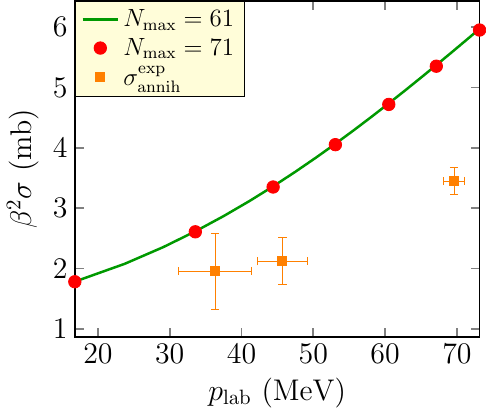}
\caption{Convergence of the $\bar{p}$-$d$ reaction cross section with respect to $N_{\max}$ (green curve and red symbols). We multiply the reaction cross section by the square of the antiproton beam's velocity in the laboratory frame, denoted with $\beta$. The cross section includes contributions from $J=1/2, \, 3/2, \,\text{and}$ $5/2$ channels of both parities. We use the deuteron ground state plus two pseudostates in the ${}^3{S}_1$-${}^3{D}_1$ channel, together with regulator parameters $r_{\text{reg}}=12$ fm and $r_{\text{reg,\text{c}}}=5$ fm. Experimental data are taken from Ref.~\cite{zenoni1999pd}. 
}
\label{fig:sigma_reac_pbar_d}
\end{figure}

%
\begin{table*}
\centering
\footnotesize
\caption{Nuclear quasibound states of the $\bar{p}$-$d$ system (in MeV). The energies are obtained by diagonalizing the Hamiltonian in the NCSM/RGM basis [\autoref{eq:phi_nu_n}] and are equivalent to enforcing bound-state asymptotic boundary conditions via the $R$-matrix method.
}
\begin{ruledtabular}
\begin{tabular}{ccccccc}
     Channel   & \multicolumn{3}{c}{$N_{\text{max}}=70$}    & \multicolumn{3}{c}{$N_{\text{max}}=80$}   \\   
                                                        & $d~ {\rm (g.s.)}$ & $d+d^*$   &   $d+2d^*$ & $d~ {\rm (g.s.)}$ & $d+d^*$   &  $d+2d^*$ \\ \hline   
     ${}^2S_{1/2}^- -{}^4D_{1/2}^-$                    & $-21.34-153.15i$  &   $-25.85-166.79i$  & $-25.86-166.83i$ & $-21.33-153.12i$& $-25.13-164.58i$  & $-25.14-164.60i$   \\    
     ${}^4S_{3/2}^- -{}^4D_{3/2}^- - {}^2D_{3/2}^-$    & $-26.55-152.64i$& $-31.29-166.16i$  & $-31.31-166.19i$ & $-26.54-152.61i$& $-30.54-163.96i$  & $-30.55-163.98i$   \\  
\end{tabular}
\label{tab:quasi_pbard_nmax_ntar}
\end{ruledtabular} 
\end{table*}
%
In \myfigsref{fig:sigma_reac_pbar_d}{fig:sigma_domega_pbar_d}, we have shown our results for the reaction and elastic differential cross section, respectively. We include contributions from total angular momenta $J=1/2,\,3/2, \,\text{and } 5/2 $ and both parities. For both observables, we investigate the dependence of our results on $N_{\text{max}}$ by comparing two values, which yield indistinguishable results. To compare with the experimental annihilation cross sections of Ref.~\cite{zenoni1999pd}, we have multiplied the reaction cross section by the square of the antiproton beam's velocity in the laboratory frame, denoted $\beta$. Accordingly, we note some discrepancy between our results and the experiment, which can again be attributed to the \NNb interaction and/or the missing configuration in the RGM expansion. It is important to note that the reaction cross section sums all the nonelastic channels, which are not necessarily annihilation. For a rigorous comparison, one has to calculate and remove the contribution from other inelastic channels, such as the charge exchange. Our comparison here is on the premise that contribution from the channels other than annihilation is small at this energy range~\cite{aghai2018measurement}. We were unable to find experimental elastic differential cross sections below the deuteron breakup threshold for a more rigorous comparison. 
\par
%
%
%
\begin{figure}
\centering
\includegraphics[width=0.4\textwidth]{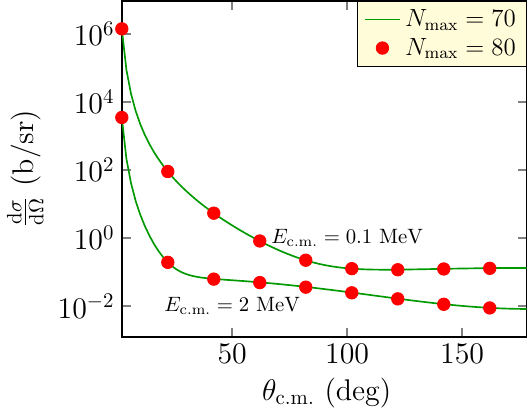}
\caption{Convergence of the $\bar{p}$-$d$ elastic differential cross section (d$\sigma$/d$\Omega$) in the center-of-mass frame with respect to $N_{\max}$ (green curve and red symbols) at $E_{\rm c.m.}=0.1$ and 2 MeV. All other parameters are the same as in \autoref{fig:sigma_reac_pbar_d}.
}
\label{fig:sigma_domega_pbar_d}
\end{figure}
\begin{figure}
\centering
\includegraphics[width=0.4\textwidth]{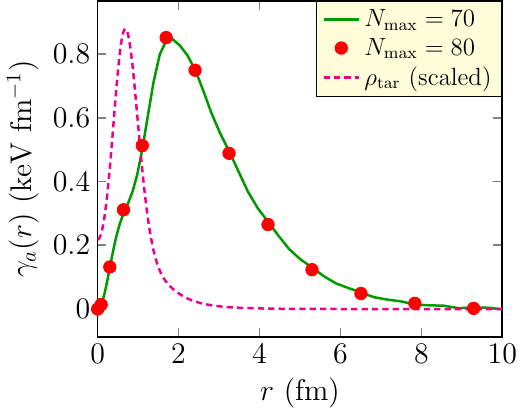}
\caption{Convergence of the $\bar{p}$-$d$ annihilation density $\gamma_a(r)$ with respect to $N_{\text{max}}$ parameter (green curve and red symbols) for the lowest atomic state in the ${}^2S^{-}_{1/2}$ channel. The dashed line shows the (scaled) deuteron matter density. Calculations include the deuteron ground state plus two pseudostates in the ${}^3{S}_1$-${}^3{D}_1$ channel and use $r_{\text{reg}}=14$ fm, $r_{\text{reg,\text{c}}}=5$ fm, $a_c=400$ fm, and $n_s=500$.
}
\label{fig:p_h2_annih_density}
\end{figure}

%
Finally, in \autoref{fig:p_h2_annih_density}, we study the convergence of the annihilation density by comparing results obtained at $N_{\rm max}=70$ and 80. The density corresponds to the lowest atomic state in the ${}^2S^{-}_{1/2}$ channel. Annihilation densities for antiprotonic deuteron were previously reported in Ref.~\cite{duerinck2023antiproton} using a Faddeev calculation with the KW \NNb potential. Our results are consistent with theirs in terms of the overall shape and the location of the maximum. Since we obtain different atomic half-widths, this difference naturally propagates into the value of $\gamma_a(r)$; however, it does not alter the qualitative discussions regarding its radial profile.
Although $\gamma_a(r)$ is not an observable, comparing it with the target matter density provides useful insight into where annihilation predominantly takes place. The target density shown in \autoref{fig:p_h2_annih_density} is computed from the Jacobi NCSM wave function following the method of Ref.~\cite{PhysRevC.70.014317}; for targets heavier than the deuteron, we employ the exact expression given in that reference. We find that, despite some penetration into the interior, the annihilation probability peaks around $r\approx 2$ fm, i.e., near the edge of the target, supporting the standard picture that low-energy antiproton annihilation is dominantly peripheral.
%
%
\subsection{Antiprotonic \texorpdfstring{${}^3$H}{3H} and \texorpdfstring{${}^3$He}{3He} systems} \label{subsec:result_a4}
We now turn to the four-body systems formed by an antiproton interacting with a three-nucleon target. As expected, the additional nucleon increases the computational cost and therefore reduces the $N_{\text{max}}$ that can be reached compared to the deuteron case. Nevertheless, we can still achieve satisfactory convergence for the observables reported below without resorting to an effective Hamiltonian. \par
\begin{figure}
\centering
\includegraphics[width=0.4\textwidth]{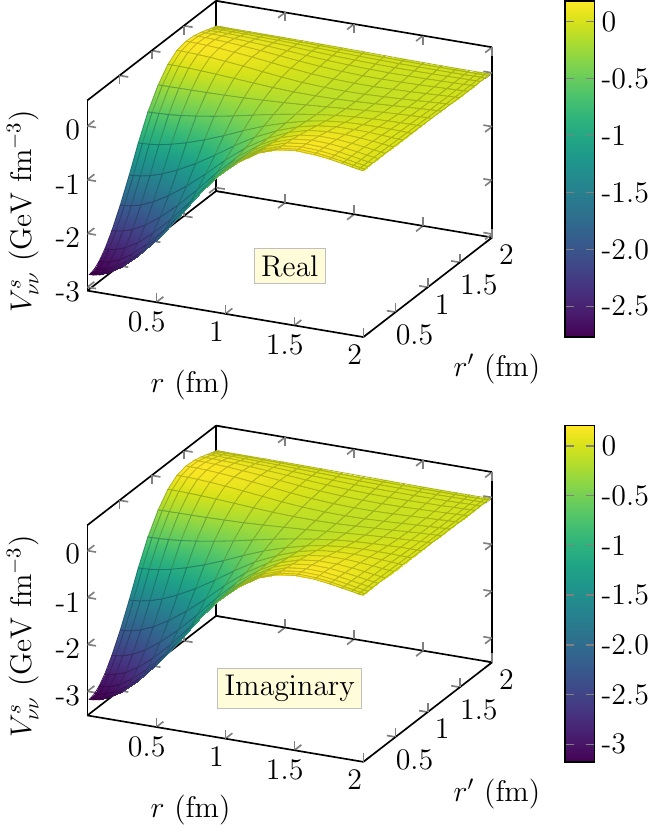}
\caption{Strong-interaction potential kernel $V^{s}_{\nu \nu}(r',r)$ for $\bar{p}$–$^3$He in the ${}^1S_{0}^-$ channel, extracted from the NCSM/RGM calculation with $N_{\text{max}}=20$ including only the target's ground state.
}
\label{fig:kernel_he3_-}
\end{figure}

We start by examining the strong-interaction potential kernels, which provide a direct point of comparison with the lighter systems discussed above. In \autoref{fig:kernel_he3_-}, we display the $s$-wave $\bar{p}$-${}^3\mathrm{He}$ kernel at $N_{\rm max}= 20$. Its overall morphology, governed by the HO radial functions, remains similar to the $\bar{p}$-$d$ case, but the attractive pocket is noticeably deeper due to the additional nucleon in the target. This stronger attraction suggests, in turn, the possible presence of more deeply bound (or more) nuclear quasibound states in the $\bar{p}$-${}^3 \mathrm{He}$ system.\par
%
\begin{figure}
\centering
\includegraphics[width=0.45\textwidth]{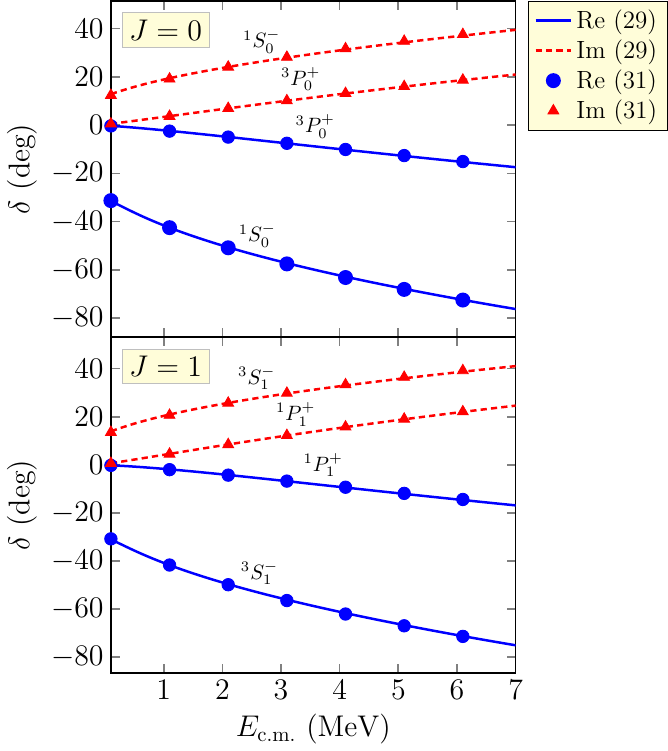}
\caption{Real (blue) and imaginary (red) diagonal phase shifts for the $\bar{p}$-${}^3 \mathrm{He}$ system computed with the NCSM/RGM including only the ${}^3$He ground state. The $N_{\text{max}}$  convergence is illustrated by comparing results at $N_{\text{max}}=28(29)$ (lines) and $N_{\max}=30(31)$ (symbols). The values in parentheses in the plot legend indicate the positive parity model space. The remaining parameters are $r_{\text{reg}}=10$ fm, and $r_{\text{reg,\text{c}}}=5$ fm. The top panel shows the $J=0$ partial waves, and the bottom panel the $J=1$ partial waves.
}
\label{fig:p_he3_phase}
\end{figure}

%
%
In \myfigsref{fig:p_he3_phase}{fig:p_h3_phase}, we present the $\bar{p}$-${}^3 \mathrm{He}$ and $\bar{p}$-${}^3 \mathrm{H}$ phase shifts, respectively. For each target, we show phase shifts from $J=0$ and $J=1$ channels, including both parities. A comparison of two calculations at $N_{\rm max}=29$ and 31 indicates satisfactory convergence.
The differences between the ${}^3$He and ${}^3$H results arise from Coulomb effects and from the slightly different thresholds. In addition, the isospin content is different: $\bar{p}$-${}^3 \mathrm{He}$ couples to both $T=0$ and $T=1$ components, whereas $\bar{p}$-${}^3 \mathrm{H}$ is restricted to $T=1$. We also observe that the phase shifts are largely governed by the relative orbital angular momentum: $s$- and $p$-wave channels with different total angular momentum and channel spin exhibit similar magnitudes and trends.
Overall, the real parts of the phase shifts are repulsive in all channels considered, and we find no evidence for low-energy resonant behavior in the energy regime studied, where nuclear breakup can be safely neglected. The dependence of these phase shifts on the regulator parameters is discussed in Ref.~\cite{dehghani}. \par
The $N_{\text{max}}$ dependence of the ${}^1S^-_0$ and ${}^3S^-_1$ scattering lengths of the ${}^3$H and ${}^3$He antiprotonic systems is summarized in \autoref{tab:p_triton_he3_scat_both}. As in the antiprotonic deuteron case, we are able to reach large model spaces for these systems, and the resulting scattering lengths show no visible dependence on $N_{\text{max}}$. These two $s$-wave channels have similar values for scattering length. Furthermore, the $s$-wave scattering length results of $\bar{p}$-${}^3 \mathrm{He}$ and $\bar{p}$-${}^3 \mathrm{H}$ are close to each other. Both real and imaginary parts are smaller in magnitude than $s$-wave values for deuteron.\par
%
%
%
%
\begin{figure}
\centering
\includegraphics[width=0.45\textwidth]{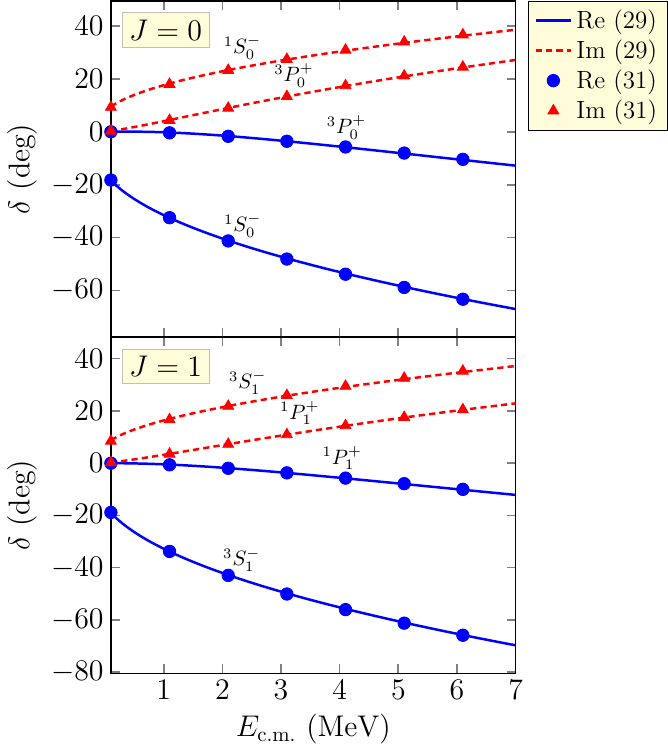}
\caption{Real (blue) and imaginary (red) parts of the $\bar{p}$-${}^3 \mathrm{H}$ scattering phase shifts, computed with the same setup and parameters as in \autoref{fig:p_he3_phase}.
}
\label{fig:p_h3_phase}
\end{figure} 
%
%
%
All the results discussed so far were obtained with the target ground state only. To quantify the impact of additional target eigenstates, \autoref{tab:p_h3_he3_scat_both_gs+1} reports calculations including excited unbound target states. For both ${}^3$He and ${}^3$H, the first pseudostate lies close to---but slightly below---the dissociation threshold; it should nevertheless be viewed as a discretized continuum state associated with target breakup, whose threshold corresponds to an excitation energy of $\approx 5$ MeV.
The scattering lengths are extracted from the ERE around $E_{\text{c.m.}}\approx 0.1$ MeV and extrapolated to $E_{\text{c.m.}}\to 0$. Since these pseudostate channels remain somewhat remote (and closed) in the low-energy regime considered here, their impact on the scattering observables is limited, in line with the discussion in \autoref{subsec:result_a3}. In the present case, including the first pseudostate changes the scattering lengths by at most $\approx 3\%$ in the real part and $\approx 10\%$ in the imaginary part. Given this modest impact and the increased computational cost, we compute the remaining observables using the ground state only. We will see below that any discrepancy with respect to exact few-body calculations cannot be removed solely by including extra target pseudostates, indicating that missing configurations are involved. \par
%
\begin{table}
\caption{$\bar{p}$-${}^3 \mathrm{H}$ and $\bar{p}$-${}^3 \mathrm{He}$ $s$-wave scattering lengths (in fm) as a function of $N_{\text{max}}$ parameter, computed using only the target's ground state.
}
\small
\centering
\begin{ruledtabular}
\begin{tabular}{cccc}
\multicolumn{4}{c}{${}^3$H} \\
\midrule
\multicolumn{2}{c}{${}^1S_0^-$} & \multicolumn{2}{c}{${}^3S_1^-$} \\
\midrule
$N_{\text{max}}=44$ & $N_{\text{max}}=48$ & $N_{\text{max}}=40$ & $N_{\text{max}}=44$ \\
$1.78-0.93i$        & $1.78-0.93i$        & $1.85-0.85i$        & $1.85-0.85i$ \\  \bottomrule 
\multicolumn{4}{c}{${}^3$He} \\
\midrule
\multicolumn{2}{c}{${}^1S_0^-$} & \multicolumn{2}{c}{${}^3S_1^-$} \\
\midrule 
$N_{\text{max}}=48$ & $N_{\text{max}}=52$ & $N_{\text{max}}=40$ & $N_{\text{max}}=44$ \\
$1.72-0.83i$        & $1.72-0.83i$        & $1.67-0.90i$        & $1.67-0.90i$  \\ 
\end{tabular}
\end{ruledtabular}
\label{tab:p_triton_he3_scat_both}
\end{table}
%
%
\begin{table}
\caption{$\bar{p}$-${}^3 \mathrm{H}$ and $\bar{p}$-${}^3 \mathrm{He}$ $s$-wave scattering lengths (in fm) as a function of $N_{\text{max}}$ parameter, computed using the target ground state plus the first pseudostate (excited state) in the same channel. 
} 
\small
\centering
\begin{ruledtabular}
\begin{tabular}{cccc}
\multicolumn{4}{c}{${}^3$H} \\
\midrule
\multicolumn{2}{c}{${}^1S_0^-$} & \multicolumn{2}{c}{${}^3S_1^-$} \\
\midrule 
$N_{\text{max}}=42$ & $N_{\text{max}}=44$ & $N_{\text{max}}=38$ & $N_{\text{max}}=40$ \\
$1.82-0.87i$        & $1.82-0.87i$        & $1.89-0.78i$        & $1.89-0.78i$ \\ 
\bottomrule
\multicolumn{4}{c}{${}^3$He} \\
\midrule
\multicolumn{2}{c}{${}^1S_0^-$} & \multicolumn{2}{c}{${}^3S_1^-$} \\
\midrule 
$N_{\text{max}}=42$ & $N_{\text{max}}=44$ & $N_{\text{max}}=36$ & $N_{\text{max}}=38$ \\
$1.77-0.77i$        & $1.78-0.78i$        & $1.72-0.82i$        & $1.72-0.82i$  \\
\end{tabular}
\end{ruledtabular}
\label{tab:p_h3_he3_scat_both_gs+1}
\end{table}
%
In \autoref{tab:level_a4_rmat} and \autoref{tab:p_h3_he3_level_both_true}, we show the level shifts and half-widths of antiprotonic atoms obtained using bound-state $R$-matrix calculation and the Trueman formula, respectively. For antiprotonic ${}^3$H, the $R$-matrix and Trueman methods give almost identical values. However, in the antiprotonic ${}^3$He case, the Trueman results of the lowest atomic state have a maximum error of $3\%$ (real part) and $6\%$ (imaginary part). This is expected because the expansion parameter in the Trueman formula is inversely related to the Bohr radius and hence larger for ${}^3$He. Correspondingly, the ${}^3$He level shift is substantially larger than that of ${}^3$H, reflecting the smaller ground-state atomic radius ($\approx 19$ fm) for antiprotonic ${}^3$He. In \autoref{tab:p_h3_he3_level_both_true_compar}, we have compared our level shifts and half-widths with the \textit{ab initio} results of Ref.~\cite{Duerinck:2026otx}, which are obtained using the Faddeev method with the second-order Trueman formula, and the same $NN$ and $N\bar{N}$ interaction as ours. For higher accuracy, our results are calculated using the scattering lengths of \autoref{tab:p_h3_he3_scat_both_gs+1}, which includes a pseudostate of the target in the calculation. A comparison of the real and imaginary parts between the two methods shows that we have a deviation between 2 and $15\%$ for the helium target and between 4 and $23\%$ for triton.\par 
%
\begin{table}
\caption{$\bar{p}$-${}^3 \mathrm{H}$ and $\bar{p}$-${}^3 \mathrm{He}$ $s$-wave antiprotonic-atom level shifts and half-widths (in keV) as a function of $N_{\text{max}}$ parameter, computed with the bound-state $R$-matrix method using the target's ground state. Other parameters are $a_c=400$ fm, and $n_s=300$.
}
\small
\centering
\begin{ruledtabular}
\begin{tabular}{ccccc}
\multicolumn{5}{c}{${}^3$H} \\
\midrule 
    & \multicolumn{2}{c}{${}^1S_0^-$} & \multicolumn{2}{c}{${}^3S_1^-$} \\
\midrule 
$n$ &$N_{\text{max}}=44$    & $N_{\text{max}}=48$   & $N_{\text{max}}=44$   &   $N_{\text{max}}=46$ \\
1   & $3.1-1.27i$           & $3.1-1.27i$           & $3.16-1.14i$          & $3.16-1.14i$ \\ 
2   & $0.41-0.18i$          &  $0.41-0.18i$         & $0.42-0.16i$          & $0.42-0.16i$ \\
\bottomrule
\multicolumn{5}{c}{${}^3$He} \\
\midrule  
    & \multicolumn{2}{c}{${}^1S_0^-$} & \multicolumn{2}{c}{${}^3S_1^-$} \\
\midrule 
$n$ & $N_{\text{max}}=48$   & $N_{\text{max}}=52$ & $N_{\text{max}}=44$ & $N_{\text{max}}=46$\\
1   & $20.6-6.0i$           & $20.6-6.0i$         & $20.48-6.6i$        & $20.47-6.6i$\\ 
2   & $2.87-0.96i$          & $2.87-0.96i$        & $2.84-1.05i$        & $2.84-1.05i$ \\
\end{tabular}
\end{ruledtabular}
\label{tab:level_a4_rmat}
\end{table}
%
%
\begin{table}
\caption{$\bar{p}$-${}^3 \mathrm{H}$ and $\bar{p}$-${}^3 \mathrm{He}$ $s$-wave antiprotonic-atom level shifts and half-widths (in keV), computed using the Trueman formula up to the second order. The calculations use the scattering lengths of \autoref{tab:p_triton_he3_scat_both}.
}
\small
\centering
\begin{ruledtabular}
\begin{tabular}{ccccc}
   & \multicolumn{2}{c}{${}^3$H}   & \multicolumn{2}{c}{${}^3$He} \\
\midrule 
$n$& ${}^1S_0^-$ & ${}^3S_1^-$     & ${}^1S_0^-$   & ${}^3S_1^-$ \\
1  &$3.11-1.29i$ & $3.18-1.16i$    & $21.05-5.64i$ & $21.01-6.35i$ \\ 
2  &$0.41-0.19i$ & $0.43-0.18i$    & $3.0-1.17i$   & $2.95-1.29i$ \\
\end{tabular}
\end{ruledtabular}
\label{tab:p_h3_he3_level_both_true}
\end{table}
\begin{table}
\caption{Comparison of our $\bar{p}-^3$H and $\bar{p}-{}^3$He $s$-wave antiprotonic-atom level shifts and half-widths with the theoretical results of Ref.~\cite{Duerinck:2026otx}, which is obtained using the Faddeev method. The results of both methods are obtained using the Trueman formula up to the second order, and correspond to the ground state of the antiprotonic atom. Our results use the scattering lengths of \autoref{tab:p_h3_he3_scat_both_gs+1} as input.
}
\scriptsize
\centering
\begin{ruledtabular}
\begin{tabular}{ccccc}
   & \multicolumn{2}{c}{${}^3$H}   & \multicolumn{2}{c}{${}^3$He} \\
\midrule 
 & ${}^1S_0^-$ & ${}^3S_1^-$     & ${}^1S_0^-$   & ${}^3S_1^-$ \\
NCSM/RGM  &$3.14-1.19i$ & $3.22-1.05i$    & $21.24-5.06i$ & $21.01-5.57i$ \\ 
Faddeev  &$2.57-1.15i$ & $2.78-0.94i$    & $18.6-4.87i$   & $18.4-5.68i$ \\
\end{tabular}
\end{ruledtabular}
\label{tab:p_h3_he3_level_both_true_compar}
\end{table}
%
%
\par
Nuclear quasibound states found in these systems are shown in \autoref{tab:p_he3_triton_quasi} for antiprotonic ${}^3$H and ${}^3$He. As before, convergence with respect to $N_{\text{max}}$ is satisfactory, and consistency with the $R$-matrix results is verified. We observe a considerable difference in the quasibound state energies of ${}^3$H and ${}^3$He targets in the ${}^1S^-_0$ channel, while those of the ${}^3SD^-_1$ channel are quite close. We emphasize that quasibound state energies presented here and those in the previous section do not have practical value in the sense that they cannot be observed in experiment due to their extremely short lifetimes. Their existence is strongly dependent on the \NNb interaction model used to describe the antiprotonic systems, a sensitivity that we do not investigate in the present work. However, if one of these states lies close to a physical threshold, it can have a significant impact on observables~\cite{Loiseau:2020cji}. It may also matter for heavier systems whose wave functions contain cluster configurations involving ${}^{3}$H or ${}^{3}$He. For this reason, it is useful to report our results here.
%
\begin{table}
\caption{Nuclear quasibound state energies of the $\bar{p}$-${}^3 \mathrm{H}$ and $\bar{p}$-${}^3 \mathrm{He}$ systems (in MeV) in the $J=0$ and $J=1$ channels, calculated using the target's ground state. The dependence on the model-space size $N_{\text{max}}$ illustrates convergence.  
}
\footnotesize
\begin{ruledtabular}
\begin{tabular}{ccc}
\multicolumn{3}{c}{${}^3$H} \\
\hline
Channel                 &  $N_{\text{max}}=40$         &  $N_{\text{max}}=44$ \\ 
${}^1S_0^-$             &  $-51.80-219.17i$            &  $-51.81-219.15i$ \\
${}^3SD_1^-$            &  $-82.06-227.51i$            &  $-82.05-227.47i$ \\
\hline
\hline
\multicolumn{3}{c}{${}^3$He} \\
\hline
                        &  $N_{\text{max}}=34(35)$   &  $N_{\text{max}}=36(37)$ \\    
${}^1S_0^-$             & $-108.78-226.95i$          & $-108.69-226.85i$ \\
${}^3SD_1^-$            & $-85.56-219.00i$           & $-85.51-218.88i$ \\
${}^1P_1^+-{}^3P_1^+$   & $-8.33-161.85i$            & $-8.25-161.67i$ \\ 
\end{tabular}
\end{ruledtabular}
\label{tab:p_he3_triton_quasi}  
\end{table}
%
\par
\begin{figure}
\centering
\includegraphics[width=0.39\textwidth]{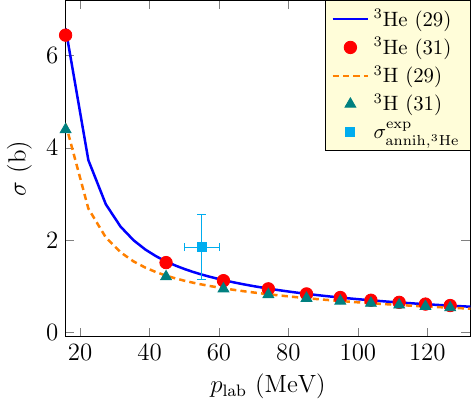}
\caption{$\bar{p}$-${}^{3}\mathrm{He}$ (blue and red)  and $\bar{p}$-${}^{3}\mathrm{H}$ (orange and green) reaction cross sections compared with experimental data. Convergence with respect to the model-space size is illustrated by results at $N_{\text{max}}=28(29)$ (lines) and $N_{\max}=30(31)$ (symbols). The results are obtained using the target's ground state, together with  $r_{\text{reg}}=10$ fm and $r_{\text{reg,\text{c}}}=5$ fm. The experimental point is taken from Ref.~\cite{bianconi2000antiproton}.
}
\label{fig:p_h3_he3_sigma_reac}
\end{figure}

We now turn to scattering observables that could directly be accessible in the current experimental facility at CERN. In \autoref{fig:p_h3_he3_sigma_reac}, we present the $\bar{p}$-${}^{3}\mathrm{H}$ and $\bar{p}$-${}^{3}\mathrm{He}$ reaction cross sections, obtained by summing contributions from $J=0,\,1$ and 2 partial waves of both parities. The results are shown as a function of the antiproton laboratory momentum to facilitate comparison with the ${}^3$He measurement of Ref.~\cite{bianconi2000antiproton}. The predicted cross sections are stable with respect to $N_{\text{max}}$, and the available experimental point lies close to our calculation.
Within the same framework, other measurable scattering observables can be computed straightforwardly (e.g., elastic angular distributions or channel-resolved inelastic cross sections or polarization observables). Measurements of these observables would help us to clarify the role of missing configurations in the RGM expansion and to discriminate between competing \NNb interaction models. At present, however, low-energy data for light antiprotonic nuclei remain scarce precisely in the regime where nuclear structure, few-body dynamics, and \NNb low-energy details are most strongly probed. More precise and systematic measurements at low energies would significantly improve constraints on the \NNb interaction and could ultimately enable reliable predictions for related processes, such as low-energy $\bar n$ scattering on the same targets, where direct data would be particularly challenging to obtain. \par
%
\begin{figure}
\centering
\includegraphics[width=0.37\textwidth]{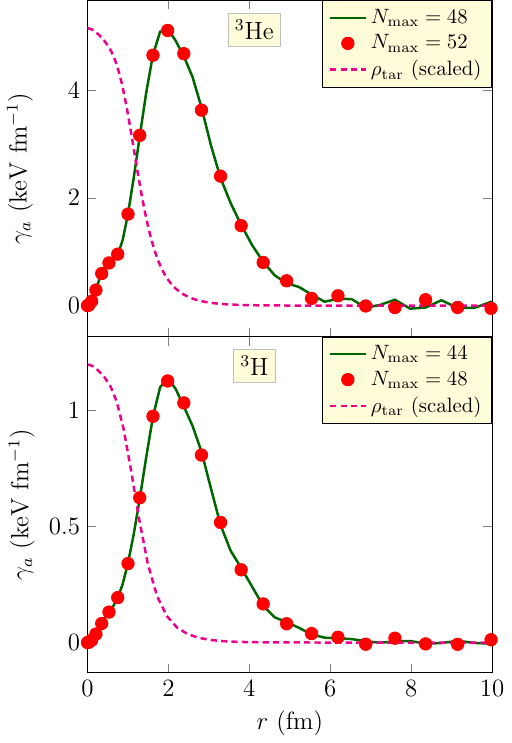}
\caption{Comparison of the $\bar{p}$-${}^{3}\mathrm{He}$ (top) and $\bar{p}$-${}^{3}\mathrm{H}$ (bottom) annihilation densities $\gamma_a(r)$ with the corresponding (scaled) $^3$He (top) and ${}^3$H matter densities (dashed lines). Convergence with respect to the model-space size $N_{\text{max}}$ is illustrated (solid green curve and symbols) for the lowest atomic state in the ${}^1S^{-}_{0}$ channel. The results are obtained using the target's ground state, $r_{\text{reg}}=10$ fm, $r_{\text{reg,\text{c}}}=5$ fm, $a_c=300$ fm, and $n_s=400$.
}
\label{fig:p_he3_annih_density}
\end{figure}

%
To address the requirements of the PUMA experiment, we wish to verify that antiproton annihilation is primarily sensitive to the tail (surface region) of the nuclear wave function. The deuteron is a somewhat special case: Its small binding energy makes it closer to a halolike system than to a typical well-bound nucleus, characterized by $\approx 8$ MeV per nucleon and a near-saturation central density. The $A=3$ targets already move us closer to this ``normal nucleus'' limit.
In \autoref{fig:p_he3_annih_density}, we therefore plot the annihilation density $\gamma_a(r)$ for the lowest atomic state in antiprotonic ${}^3$He (top) and ${}^3$H (bottom) and examine its dependence on the model-space size. For reference, the corresponding target matter density is shown in each panel. As for the deuteron, the annihilation density peaks around $r\approx 2$ fm, i.e., near the tail of the nuclear density, indicating that annihilation remains dominantly peripheral in these light systems. Moreover, our $\gamma_a(r)$ match those reported in Ref.~\cite{Duerinck:2026otx}. This agreement suggests that, despite the fact that we retain only the $\bar p$+nucleus configuration and neglect additional rearrangement configurations, we reproduce the main features of the exact few-body description within the uncertainties already quantified in the extraction of atomic levels.
We also confirm that $\gamma_a(r)$ is essentially insensitive to $N_{\rm max}$ as illustrated by the comparison of two model spaces in \autoref{fig:p_he3_annih_density}. Overall, for these light targets, the working hypothesis that annihilation occurs predominantly at the nuclear periphery appears robust. This peripherality is precisely what makes antiprotonic-atom measurements promising as indirect probes of nuclear density moments, potentially allowing one to connect annihilation observables to the proton and neutron density distributions, especially if annihilation yields from excited atomic states (e.g., the second level) can be measured.
%
%
\section{Conclusion and outlook} \label{sec:conclusion}
In this work, we applied the NCSM/RGM to study light antiprotonic systems at low energies. We presented the Jacobi NCSM/RGM formalism for antiproton-nucleus systems in detail. In these systems, no antisymmetrization is needed between the target constituents and the projectile. Consequently, the NCSM/RGM formalism becomes considerably simpler compared to nucleon-nucleus systems: No exchange contributions show up, and the norm kernel is trivial. This simplification is numerically significant and, in particular, facilitates high-accuracy implementations in the Jacobi NCSM/RGM. Consequently, we were able to perform large-model-space calculations for a wide range of observables, including scattering phase shifts, scattering length, atomic level shifts and half-widths, nuclear quasibound states, and elastic differential and reaction cross sections. We compared our results with the Faddeev method to assess the uncertainty of our many-body method for these light systems. We expect this uncertainty to shrink as we move towards heavier targets.  \par
Overall, our findings show that the NCSM/RGM can be relied on to study antinucleon-nucleus systems. One of the complications that we encountered was the numerical artifacts arising due to the introduction of ``hard'' \NNb interaction into the finite HO expansion of the NCSM/RGM kernels. We used regulators to suppress these finite-model-space artifacts. Alternatively, one can avoid this issue by taming the interaction using techniques like the similarity renormalization group (SRG)~\cite{bogner2007similarity,bogner2008three} or by using ``soft,'' bare interactions. With softer interactions, our calculations can be extended to heavier targets by using the Slater-determinant version of the method. This will be the subject of future work. \par 
In light of future proposals for experiments at CERN, calculations
with the antiproton projectile can be generalized to antideuteronic atoms by modifying the NCSM/RGM
formalism for deuteron projectiles~\cite{Navratil:2011ay}. Additionally, these developments can be applied to systems containing hyperons. In fact, we expect the NCSM/RGM formalism for
these systems to share some similarities with its antiprotonic counterpart.
\begin{acknowledgments}
AD and GH express their gratitude to Jaume Carbonell, S\l awomir Wycech, Pierre-Yves Duerinck, and Rimantas Lazauskas for insightful discussions and for providing benchmarks. This material is based in part upon work supported by the U.S. Department of Energy, Office of Science, Office of Nuclear Physics, under Work Proposal No. SCW0498. This work was prepared in part by LLNL under Contract No. DE-AC52-07NA27344. PN acknowledges support from the NSERC Grant No. SAPIN-2022-00019. TRIUMF receives federal funding via a contribution agreement with the National Research Council of Canada. This project was provided with computing HPC and storage resources by GENCI at IDRIS/TGCC, thanks to the Grant No. 2015-0513012 on the supercomputer Jean Zay/Joliot Curie. GH gratefully acknowledges support from the CNRS/IN2P3 Computing Center (Lyon - France) for providing computing and data-processing resources needed for this work. GH and AD acknowledge the ANR-FRANCE (French National Research Agency) for its financial support (Grant No. ANR-21-CE31-0020).
\end{acknowledgments}
\bibliography{references}
\end{document}